\documentclass[10pt,a4paper,twoside,twocolumn,american,prd,nofootinbib,superscriptaddress]{revtex4}
\usepackage{lmodern}

\usepackage[T1]{fontenc}
\usepackage[utf8]{inputenc}
\usepackage{graphicx}
\usepackage{color}
\usepackage{babel}
\usepackage{natbib}
\usepackage{booktabs}
\usepackage{units}
\usepackage{amsmath}
\usepackage{amssymb}
\usepackage{esint}
\usepackage[normalem]{ulem}
\usepackage[unicode=true,pdfusetitle,
 bookmarks=true,bookmarksnumbered=false,bookmarksopen=false,
 breaklinks=false,pdfborder={0 0 0},backref=false,colorlinks=true]
 {hyperref}
\hypersetup{
 citecolor=blue,filecolor=blue,linkcolor=blue,urlcolor=blue}
\usepackage{subfigure} 
\usepackage{float}
\usepackage{verbatim}



\def\kpc{{\rm kpc}}
\def\mpc{{\rm Mpc}}
\def\rc{{r_{\rm cut}}}

\def\vir{{\rm vir}}

\begin{document}

\title{3D simulations with boosted primordial power spectra and ultracompact minihalos}

\author{Mateja Gosenca}
\email{M.Gosenca@sussex.ac.uk}
\affiliation{Astronomy Centre, School of Mathematical and Physical Sciences, University of Sussex, Brighton BN1 9QH, United Kingdom}

\author{Julian Adamek}
\email{Julian.Adamek@obspm.fr}
\affiliation{Laboratoire Univers et Th\'eories, Observatoire de Paris -- PSL Research University -- CNRS -- Universit\'e Paris Diderot -- Sorbonne Paris Cit\'e,
5 place Jules Janssen, 92195 Meudon CEDEX, France}

\author{Christian T.~Byrnes}
\email{C.Byrnes@sussex.ac.uk}
\affiliation{Astronomy Centre, School of Mathematical and Physical Sciences, University of Sussex, Brighton BN1 9QH, United Kingdom}

\author{Shaun Hotchkiss}
\email{S.Hotchkiss@auckland.ac.nz}
\affiliation{Astronomy Centre, School of Mathematical and Physical Sciences, University of Sussex, Brighton BN1 9QH, United Kingdom}
\affiliation{Department of Physics, University of Auckland, Private Bag 92019, Auckland, New Zealand}

\begin{abstract}
We perform three-dimensional simulations of structure formation in the early Universe, when boosting the primordial power spectrum on 
$\sim \kpc$ scales. We demonstrate that our simulations are capable of producing power-law profiles close to the steep $\rho\propto r^{-9/4}$ halo profiles 
that are commonly assumed to be a good approximation to ultracompact minihalos (UCMHs). However, we show that for
more realistic initial conditions in which halos are neither perfectly symmetric nor isolated, the steep power-law profile is
disrupted and we find that the Navarro-Frenk-White profile is a better fit to most halos. In the presence of background fluctuations even extreme, nearly spherical initial conditions do not remain exceptional. 
Nonetheless, boosting the amplitude of initial fluctuations causes all structures to form earlier and thus at larger densities. With sufficiently large amplitude of fluctuations we find 
that values for the
concentration of typical halos in our simulations can become very large.
However, despite the signal coming from dark matter annihilation inside the cores of these halos
being enhanced, it is still 
orders-of-magnitude smaller compared to the usually assumed UCMH profile. The upper bound on the primordial power spectrum from the
non-observation of UCMHs should therefore be re-evaluated.
\end{abstract}
\maketitle

\section{Introduction}

The cosmic microwave background (CMB) has yielded precise constraints on the primordial power spectrum of density perturbations
that gave rise to all of the gravitationally bound structures observed in the Universe today \cite{Ade:2015xua}. These constraints are especially
tight over about two orders of magnitude in scales, with the constraints on the largest scales being weakened by cosmic variance limits, which cannot be overcome. Hence the search for new information about the primordial power spectrum must focus on the smaller scale perturbations where there is potentially enormously more information to gain. 
There are numerous challenges to measuring the power spectrum on scales $k~\gtrsim 0.3~h\mpc^{-1}$, e.g. Silk damping of the CMB, radiation pressure during structure formation, and the nonlinear dynamics of matter in the late Universe.

A promising constraint comes from
the non-detection of primordial black holes (PBHs), which puts upper limits on the primordial power spectrum over a vast range of
scales spanning 20 orders of magnitude \cite{Carr:2009jm,Josan:2010cj}. However,
the constraints are weak, around ${\mathcal P}^\zeta\lesssim 10^{-3}-10^{-2}$, six orders of
magnitude above the amplitude observed on CMB scales. 
While these upper limits do provide useful constraints on certain models of
inflation \cite{Kawasaki:2016pql,Pattison:2017mbe}, they are very weak because the critical threshold
to collapse to a PBH is extremely large and only a curvature perturbation of order unity will do so (with an exception if there was an early
matter-dominated era, during which time the critical collapse threshold is greatly reduced by the absence of pressure \cite{Carr:2017edp,Cole:2017gle}). 

If the initial perturbations had a higher amplitude than expected from extrapolating the CMB measurement then another possible signature would be ultracompact minihalos (UCMHs), which are dense dark
matter halos with a very steep density profile and large central density. It has been claimed that the non-detection of annihilating
dark matter signatures from such compact objects provides relatively stringent upper limits on the primordial power spectrum, around
${\mathcal P}^\zeta\lesssim 10^{-7}-10^{-6}$ over the scales corresponding to $k$ from $10\,\mpc^{-1}$ to $10^7\,\mpc^{-1}$
\cite{Bringmann:2011ut}.
Due to the exponential
dependence of the number density of UCMHs and PBHs on the amplitude of the power spectrum, confirmation of the UCMH constraints would
also completely rule out the existence of any PBHs forming on these scales (corresponding to a PBH mass greater than about a solar mass). This would rule out the observed supermassive black holes having a primordial origin \cite{Kohri:2014lza}. The relation between PBH and UCMH formation has also been studied in \cite{Mack:2006gz,Lacki:2010zf,Saito:2010ts}.

Assuming that weakly interacting massive particles (WIMPs) make up the majority of dark matter, the most stringent observational constraint on the existence of UCMHs comes from the 
expected annihilation signal from the cores of UCMHs, 
where the density is extremely large and hence the probability of WIMPs annihilating 
is massively boosted by their large number density. The constraints only weakly depend on the assumed cross section and mass of the WIMPs,
but can 
be totally evaded if dark matter does not annihilate.
There are nonetheless other possible signatures of UCMHs, 
such as from the Shapiro time-delay of millisecond pulsars (although note that no actual constraints exist yet, but there are forecasts \cite{Clark:2015sha, Clark:2015tha}) or microlensing \cite{Ricotti:2009bs,Zackrisson:2012fa,Li:2012qha}.

It is in any case interesting to ask how the dark matter substructure would change if the initial small scale perturbations were larger than expected.
Other papers which constrain the power spectrum on small scales include \cite{Zhang:2010cj,vandenAarssen:2012ag,Chluba:2012we,Yang:2012qi,Yang:2012iw,Shandera:2012ke,Yang:2013dsa,Berezinsky:2013fxa,Ben-Dayan:2013eza,Berezinsky:2014wya,Yang:2015gua,Natarajan:2015cva,Anthonisen:2015tda,Aslanyan:2015hmi,Clark:2016pgn,Beck:2016gkv,Emami:2017fiy,Choi:2017ncz}. In particular, we note that CMB spectral distortion constraints provide an upper bound on the primordial power spectrum on about the same scales as UCMHs, which are about an order of magnitude weaker than the currently claimed constraints but independent of the DM model \cite{Chluba:2012we}.
 
To date, the theoretical forecasts on UCMH formation have been made using an idealised analytical model of their formation. Josan
and Green \cite{Josan:2010vn} assumed that any density contrast $\delta=\delta\rho/\bar{\rho}\,$ satisfying $\delta_c>10^{-3}$ at horizon entry would form an
UCMH (following \cite{Ricotti:2009bs}) with a steep power law profile $\rho\propto r^{-9/4}$. This is motivated by the analytical calculation of Bertschinger 1985 \cite{1985ApJS...58...39B}, who calculated the profile of a spherically symmetric perturbation accreting from a homogeneous background.
The analytical calculation of the requisite density threshold was refined by Bringmann et al.\ \cite{Bringmann:2011ut} who included the effects of radiation in 
their calculation, which is relevant because UCMHs are assumed to form shortly after matter-radiation equality. They found comparably tight 
constraints. We show that in practice no $\delta_c$ actually exists, because the formation and final profile of a small halo strongly 
depends on its environment in a way which cannot be captured even approximately by a single number, unlike the case for PBHs.

In this paper we revisit and dramatically revise the calculation of UCMH formation, and for the first time perform realistic 3D simulations 
of this complex process. The analytical calculations assumed that UCMHs are sufficiently rare high-density peaks that form in 
isolation, surrounded by an unperturbed background until redshift $\sim 10$. They also neglected the effects of angular momentum, thereby using 
spherical symmetry to reduce the problem to a 1D one. This gives rise to the steep $\rho\propto r^{-9/4}$ density profile, with a 
correspondingly boosted dark matter annihilation signal. 
We verify these results with an N-body simulation of an isolated spherical overdensity, obtaining a UCMH-like density profile.

In practice however, all UCMHs will be formed with particles with non-zero initial peculiar velocities and they are not really isolated from 
nearby perturbations, even if the nearest perturbations have a smaller amplitude. We analytically estimate that even an exceptionally large 
amplitude perturbation will be relatively close (in comparison to the size of the object) to another perturbation with the same scale and at 
least half its amplitude, even in the case of non-Gaussian initial perturbations. Moreover, even if an overdense region appears spherical when smoothed on a particular scale, it will still have fluctuations on smaller scales. These fluctuations will themselves grow, making the initial sphericity unstable \cite{2011ApJ...734..100L}.

We then perform N-body simulations using two classes of 
initial conditions: the first with a $1\, \kpc/ h$ sized protohalo of large initial density surrounded by successively larger surrounding 
fluctuations, and the second
starting with the standard featureless power law spectrum of the curvature perturbation but boosted
around the $\kpc$ scale, with boost factors 
of $10,\;100$, $1000$, 
and
without the boost (corresponding to the standard $\Lambda$CDM cosmology).

We find 
no clear evidence that 
UCMHs are formed 
even in the case when we boost the initial power spectrum by a factor of 1000.
Furthermore, a UCMH which 
formed when simulated in isolation becomes unexceptional compared to other halos in the simulation box when we add background perturbations with a typical amplitude 1/5 of that of the protohalo. This means that the analytical estimates of UCMH formation are a poor match to realistic initial conditions. Instead we find that the emergence of NFW profiles is more generic \cite{Angulo:2016qof,Ogiya:2017hbr,2010arXiv1010.2539D}. NFW profiles have a central core with a much gentler $\rho\propto 1/r$ 
slope, and a correspondingly much weaker WIMP-annihilation signal.

The NFW minihalos formed when the initial power spectrum is boosted are nonetheless of 
much greater density than would exist in the standard $\Lambda$CDM Universe. Because \emph{every} structure that forms earlier forms at a greater density, these NFW minihalos are also much more abundant than proposed UCMHs, which would only have formed at the most extreme locations of the primordial density field. Therefore it remains interesting to study which observational signatures they 
may give rise to. We provide estimates of how sharply the existing constraints have to be weakened when taking our results into account by
calculating the WIMP annihilation signal from an NFW rather than the 
$r^{-9/4}$ profile. 

This paper is structured as follows: In the next section 
 we introduce the power-law density profile usually assumed for the UCMHs and the universal NFW profile, and explain the fitting procedure we used to find the parameters of these models. We also describe the initial conditions of large-amplitude fluctuations.  In section \ref{sec:spherical-overdensity} we discuss the simulation results 
from an initial spherical overdensity, whose evolution into an UCMH in a homogeneous background we disrupt by adding random perturbations around it, with successively larger amplitude. We 
then boost the power spectrum on the $\kpc$ scale in section \ref{sec:PSboost} by up to a factor of one thousand, and we describe how this gives rise to many NFW-like halos with very large concentration parameters. In section \ref{sec:WIMP} we 
briefly review
the theory of WIMP annihilation in the dense center of halos and calculate the signal from the halos we simulated, before concluding in section \ref{sec:conclusions}. Technical details of how we performed the simulations and convergence testing are left to the appendices.

\section{Halo profiles and properties}

\subsection{UCMH profiles}
Ultracompact minihalos are dark matter halos, expected to form around matter-radiation equality, featuring a very steep power-law density profile \cite{Ricotti:2009bs}
\begin{equation}
\label{eq:UCMHprofile}
\rho(r) \propto r^{-9/4} \; .
\end{equation}
If a UCMH forms, its extreme compactness might allow it to retain its shape until the present time making it and the extreme density at its center potentially observable.

This $r^{-9/4}$ power-law profile is the late-time form for the density of an initial spherical overdensity accreting from a homogenous background \cite{1985ApJS...58...39B}.\footnote{Note that this is only strictly true when the profile is sufficiently coarse-grained. The true profile will be made up of a series of separate caustics arising from shells that have passed through the center of the overdensity a different number of times. See Fig.~8 of \cite{1985ApJS...58...39B} for the shape of the full non-coarse-grained profile.} This is true irrespective of the density profile of the initial seed overdensity, so long as its size is finite, it is spherical, and it eventually starts accreting from a homogeneous background. Though note that even the smallest amount of asphericity in initial conditions is unstable and will result in a triaxial profile \cite{2011ApJ...734..100L}.

For fitting to the power-law profiles we use the ansatz: 
\begin{equation}
\rho(r) = C r^{-\alpha}.  
\end{equation}
If we know the mass within a certain radius then $C$ can be calculated via:
\begin{equation}
\label{eq:massintegral}
M(r_{\rm max}) = \int_{0}^{r_{\rm max}} 4\pi r^2 dr \rho(r). 
\end {equation}
The choice of $r_{\rm max}$ is somewhat ambiguous, because halos do not have a clear ``edge''; their density instead asymptotes to the average density of the Universe and, for $\alpha\leq 3$, the above integral diverges with $r_{\rm max}$. There are several possible choices for $r_{\rm max}$ in the literature, we use the virial radius $r_{\rm vir}$ as determined by the \textit{ROCKSTAR} halo finder  \cite{Behroozi:2011ju}. It is defined as the radius of a sphere inside which the average density contrast is $18 \pi^2\simeq 178\,$\cite{Bryan:1997dn}. Correspondingly, the mass of a halo up to the virial radius $M_{\rm vir} = M(r_{\rm vir})$ is the virial mass.

For the above power-law profile, $C$ can therefore be expressed in terms of the virial mass and virial radius as:
\begin{equation}\label{eq:powermass}
C=\frac{M_{\vir}}{4\pi}(3-\alpha) r_{\vir}^{(\alpha-3)} .
\end{equation} 

\subsection{The Navarro-Frenk-White profile}
Most of the halos in the Universe appear to exhibit a density profile close to the Navarro-Frenk-White (NFW) \citep{Navarro:1995iw, Navarro:1996gj} profile:
\begin{equation}
\label{eq:NFWprofile}
\rho(r)=\frac{\rho_0}{\frac{r}{r_s}\left(1+\frac{r}{r_s}\right)^2} \; ,
\end{equation}
characterised by two parameters: $r_s$, the ``scale radius'', and $\rho_0$. For small radii, $r \ll r_s$, the profile's radial dependence is $\rho(r) \propto 1/r$, and for large $r$, it is $\rho(r) \propto 1/r^3$. The scale radius $r_s$ therefore determines the radius at which the profile changes from one power law to the other. The characteristic density $\rho_0$ corresponds to $\rho_0 = 4\rho(r_s)$. 
Integrating Eq.~(\ref{eq:massintegral}) for $\rho(r)$,
we obtain the mass within radius $r_{\rm max}$:    
\begin{equation}
M(r_{\rm max}) = 4 \pi \rho_0 r_s^3 \left( \ln\left(\frac{r_s+r_{\rm max}}{r_s}\right)- \frac{r_{\rm max}}{r_s + r_{\rm max}}\right).
\end{equation}
Defining the concentration parameter \citep{Bullock:1999he} as:
\begin{equation}
c=\frac{r_{\rm vir}}{r_s}\; ,
\end{equation}
$\rho_0$ can be expressed as:
\begin{equation}
\label{eq:NFWmass}
\rho_0 = \frac{M_{\rm vir}}{4 \pi r_s^3 \left( \ln(1+c) -c/(1+c)\right)} \; .
\end{equation}

\subsection{Fitting the profiles}\label{sec:fitting}
We can use the relation between $M_{\vir}$, $r_{\vir}$ and the two profiles' parameters (i.e. equations \eqref{eq:powermass} and \eqref{eq:NFWmass}) to eliminate one free parameter from each profile. In logarithmic space the NFW profile can be expressed as:
\begin{equation}
\begin{split}
\log\left( \rho(r)\right) = &\log \left( \frac{M_{\vir}}{4\pi r_{\vir}^3}\right) \\ 
&+ \log \left( \frac{c^2}{\ln(1+c)-c/(1+c)}\right) \\
&-\log \left( r/r_{\vir}\right) -2\log \left(1+ c r/r_{\vir}\right) \; 
\end{split}
\end{equation}
and the power-law as 
\begin{equation}
\begin{split}
\log\left( \rho(r)\right) = &\log \left( \frac{M_{\vir}}{4\pi r_{\vir}^3}\right)\\ 
&+ \log(3-\alpha) - \alpha \log\left(r/r_{\vir} \right).
\end{split}
\end{equation}
For $M_{\vir}$ and $r_{\vir}$ we use the values calculated with \textit{ROCKSTAR}.

To perform the fit we then logarithmically weight every point in the profile. Specifically, the function we minimise when fitting to a profile is:
\begin{equation}\label{eq:fittingstat}
S = \sum_i w_i \Big(\log(\rho_{\rm model}(r_i)) - \log(\rho_{\rm data}(r_i))\Big)^2 \; ,
\end{equation}
with the weights: $w_i = \log(r_{i} / r_{i-1})$ chosen so that each point is weighted according to the logarithm of the size of the bin it represents. This would be equivalent to having bins of equal length in $\log(r)$ with equal weighting. This fitting was chosen because we fit over more than an order of magnitude in $r$ and the density within this range also changes by more than an order of magnitude.

The fitting is performed
over the range from $r=0.004\, \kpc/h$ to $r= r_\vir$, independently for each halo. The lower bound is forced upon us by the resolution of the profiles (see section \ref{sec:convergence}) and the upper bound is where we assume the `edge' of the halo to be.

For the NFW fits we can also plot rescaled density $\rho/\rho_0$, against rescaled radius $r/r_s$. In these units all halos with an NFW profile ``collapse'' onto a single curve.

One selection criterion we apply is that we only consider halos that are not in the process of merging. To determine this, we compare two measures of the halo mass. The first is the combined mass of the particles that \textit{ROCKSTAR} associates to that halo given their six-dimensional phase-space information. The second one includes all particles, even the ones whose momenta suggest that they are not part of the halo. We only keep halos where these two mass parameters agree within $10\%$.

\subsection{Extreme fluctuations}\label{sec:extreme}
To date, the literature on UCMHs has focussed on treating these halos as isolated objects, forming in an otherwise homogeneous and uniform background. This allows analytical calculations to be made but the validity of this approximation has not been verified. 

\begin{figure}[t]
\centering
    \includegraphics[width=\columnwidth]{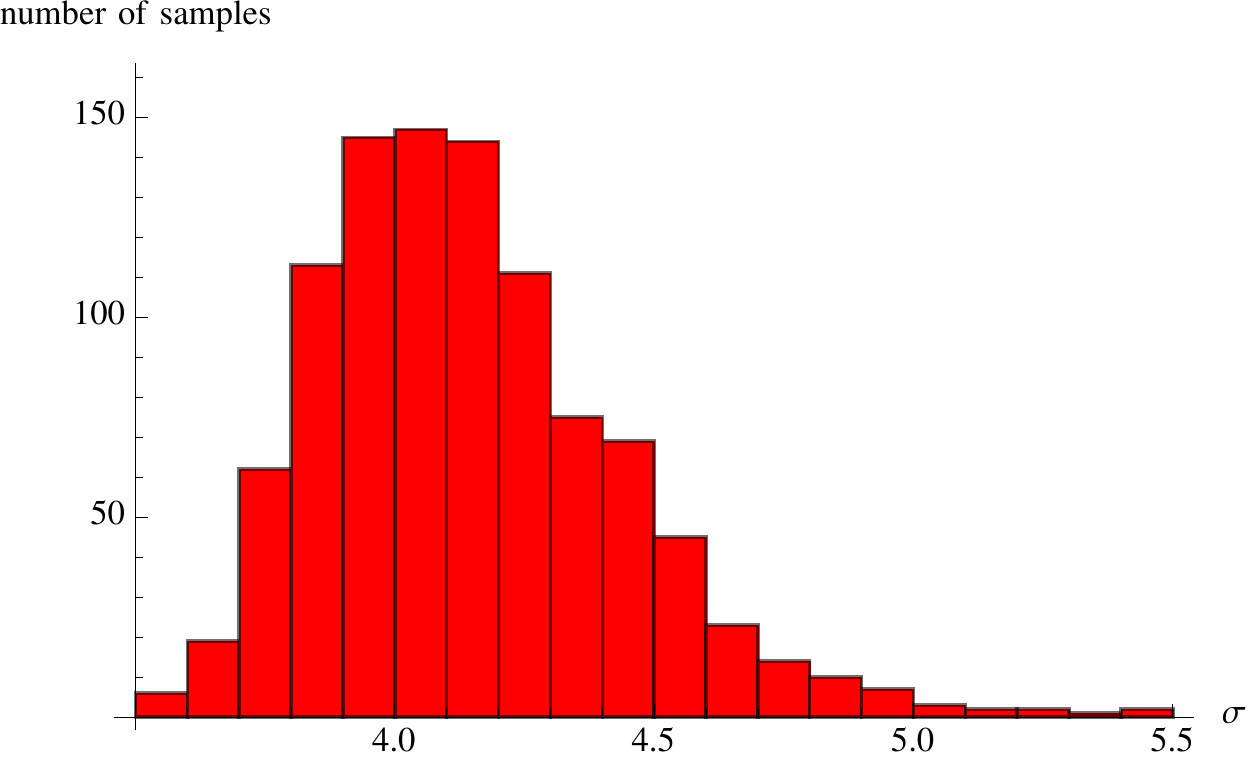}
\caption{\label{fig:extreme-value} A histogram showing how many standard deviations the maximum $\delta_{1\kpc/h}$ is away from the average in a $32 \, \kpc/h$ sized volume (taken from 1000 samples). To derive this histogram we have made the approximation that each $1\kpc/h$ region in each $32 \, \kpc/h$ volume is independent of all the other $1 \kpc/h$ regions in that volume.}
\end{figure}

To answer how extreme the most dense region of a certain size in some larger volume is, and  
assuming the density perturbation $\delta$ at horizon entry to be distributed according to the Gaussian distribution with $\bar{\delta}=0$, we can express the fraction of volume where $\delta > x $ as:
\begin{equation}
\begin{split}
\frac{V_{\delta>x}}{V} &=\frac{1}{\sigma \sqrt{2\pi}} \int_x^{\infty} e^{\left(-\frac{\delta^2}{2\sigma^2}\right)} d\delta \\
&= \frac{1}{2} \mathrm{Erfc}{\left( \frac{x}{\sqrt{2}\sigma}\right)}.
\end{split}
\end{equation}
If we further consider a region of the size of $(1 \kpc/h)^3$ in a $(32 \, \kpc/h)^3$ volume, we find that the most extreme overdensity
is a $4.0$-$\sigma$ fluctuation. We show the distribution of the most extreme values in Fig.~\ref{fig:extreme-value}, which shows that a $4$-$\sigma$ fluctuation is the most likely value, but that values as extreme as $5.5$-$\sigma$ are possible if one draws from 1000 samples.

We can also reverse the question and estimate the fraction of volume where the density is greater than some threshold value. For $2$ and $3$ times the typical density contrast, we get this fraction to be $0.023$ and $0.001$, respectively. In a $(32 \, \kpc/h)^3$ volume this corresponds to 
$745$ and $44$ regions of $(1 \kpc/h)^3$ size. 
Assuming that all peaks are randomly placed independently of each other, an estimate of the distance between the most overdense region and the one with overdensity greater or equal to twice the typical is given by
$32/(745)^{1/3}\kpc/h\simeq 3.5 \, \kpc/h$, i.e.~rather close compared to the scale of the halo itself. Due to the clustering of density peaks which arises from superimposing multiple scales, the distances between large overdensities will in fact be even smaller than the estimate made above \cite{1989MNRAS.238..293L}.

The spherical infall model, which leads to a steep power law profile also assumes spherical symmetry. Whilst it is true that the most extreme peaks are expected to be close to spherical initially \cite{Bardeen:1985tr}, the unsmoothed density field will still have substructure. This substructure will grow with time and 
pull any matter on a purely radial trajectory off this trajectory.
Moreover, gravitational tidal forces from the nearby surrounding halos will further break the spherical symmetry.

In the presence of large primordial non-Gaussianities (which are observationally unconstrained on the small scales we are interested
in), one might expect the existence of isolated large-amplitude density peaks to become much more probable.
For some probability density functions, extremely large-amplitude peaks do become
exponentially more likely, 
but so do other comparably large amplitude peaks, implying that the extremely large peak will still not be isolated. For example, if the 
density perturbation is drawn from a chi-squared probability density function with one degree of freedom, 
$\delta=\delta_G^2-\langle\delta_G^2\rangle$, where $\delta_G$ has a Gaussian distribution (which follows in the asymptotic limit of very large local $f_{\rm NL}$ \cite{Lyth:2012yp}), then the typical $(32 \, \kpc/h)^3$ volume will 
have a $10$-$\sigma$ fluctuation and the most extreme fluctuation in 1000 realisations of the initial conditions will have a $20$-$\sigma$ 
fluctuation, twice the amplitude of the second largest perturbation expected in the same volume. Furthermore, the mode coupling between 
different scales would in practice mean that the largest overdensity is likely to be situated close to the second largest perturbation, and 
so it still appears to be very hard to come up with a situation in which a large amplitude perturbation is likely to form in isolation from 
its nearest neighbours.  We will show in Sec.~\ref{sec:snr15} that a UCMH-like halo can form if we place a smooth, spherical overdensity surrounded by perturbations with a typical amplitude 15 times smaller. It would be interesting to find a model of the early Universe capable of generating comparable initial conditions. 

\section{Gaussian-profile spherical overdensity.}
\label{sec:spherical-overdensity}

\subsection{A completely isolated halo}
\begin{figure}[tb]
\centering
    \includegraphics[width=\columnwidth]{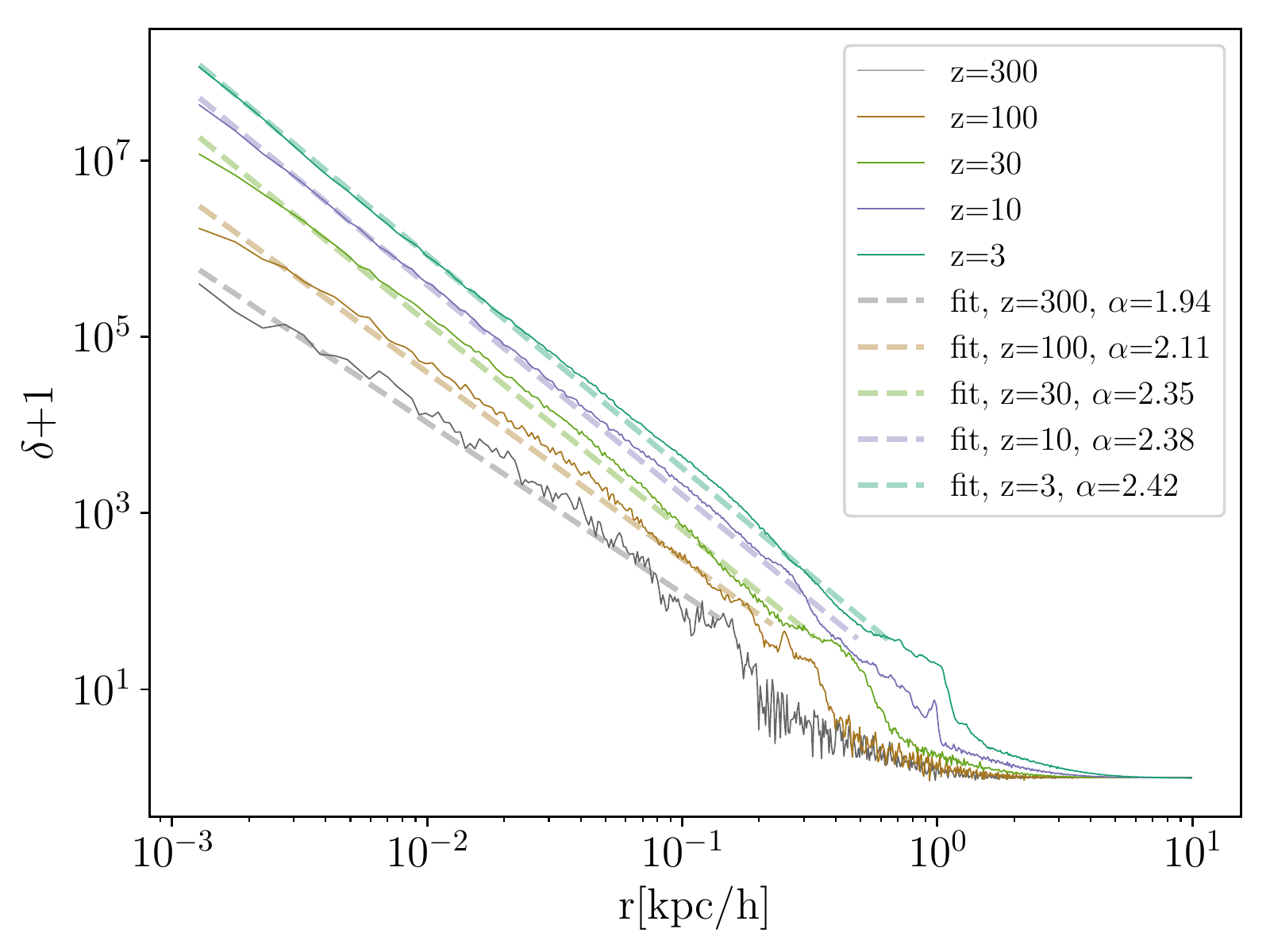} 
    \caption{\label{fig:gaussian-snrinf} (Color online) The profile of the Gaussian-seed halo in the absence of all other perturbations. We fit the power-law and thus obtain the parameter $\alpha$. The number of particles used for this simulation was $512^3$. By $z=3$ the halo mass has grown to $1.8\times 10^4 M_\odot /h$, which is still a negligible fraction of the total mass inside the box (meaning that the finite box size is not slowing down the halo accretion).}

\end{figure}

We first simulate an isolated, spherically symmetric overdensity to test the theoretical prediction of a power-law profile (\ref{eq:UCMHprofile}).
For the initial profile of the overdensity we chose a three-dimensional function with a Gaussian profile
\begin{equation}
\delta_\mathrm{in}(r)= A \exp\left(\frac{-r^2}{2\sigma_w^2}\right),
\end{equation}
where $A$ represents the initial amplitude of the seed overdensity in the center, and $\sigma_w$ regulates the size of the overdensity. We place the overdensity at the center of the box and $r$ denotes the distance from this center.
We have set $\sigma_w=0.5\, \mathrm{kpc}/h$ so that the overdensity is well-contained inside the simulation volume. 
We set $A = 0.3$ at initial redshift $z_\mathrm{in} = 10\,000$ such that the perturbation starts off marginally within the linear regime
but quickly triggers the spherical collapse process shortly after matter-radiation equality. After some time the collapse dynamics give rise to a self-similar steady-state and we determined the slope of the density profile by fitting
\begin{equation}
\label{eq:fit_fun}
\rho(r)=C r^{-\alpha},
\end{equation}
which in log-log space is just a straight line with the slope $- \alpha$. In Fig.~\ref{fig:gaussian-snrinf} we show that we indeed obtain a $\rho(r)\sim r^{-\alpha}$ profile. It is not precisely $\alpha=-9/4$, but this is not entirely unexpected because of numerical instabilities \cite{Vogelsberger:2009bn,2011ApJ...734..100L}  and the fact that $\alpha=-9/4$ is only the coarse-grained form of this type of structure growth.

Collisionless particles that can freely move through the center of the overdensity form caustics, which can be seen as additional structure at the edges of profiles. This effect is physical and we have tested that by changing the resolution of the simulation, as shown in Fig.~\ref{fig:sn15convergence}.

\subsection{Peak-to-background ratio 15}\label{sec:snr15}

Having seen that we can form a UCMH starting with a spherically symmetric overdensity surrounded by a homogeneous background, we will
now investigate how the evolution of the central halo changes when we drop the assumption of a homogeneous background. Following the
discussion in Sec.~\ref{sec:extreme}, we would not expect the amplitude of the central halo to be more than about five times larger than the 
typical amplitude of the other perturbations. We study that case in Sec.~\ref{sec:pb5}. Here we study as an intermediate case the 
situation where the amplitude of the typical fluctuations in the box smoothed on the same $\kpc/h$ scale were 15 times smaller. 

\begin{figure}
 \includegraphics[width=\columnwidth]{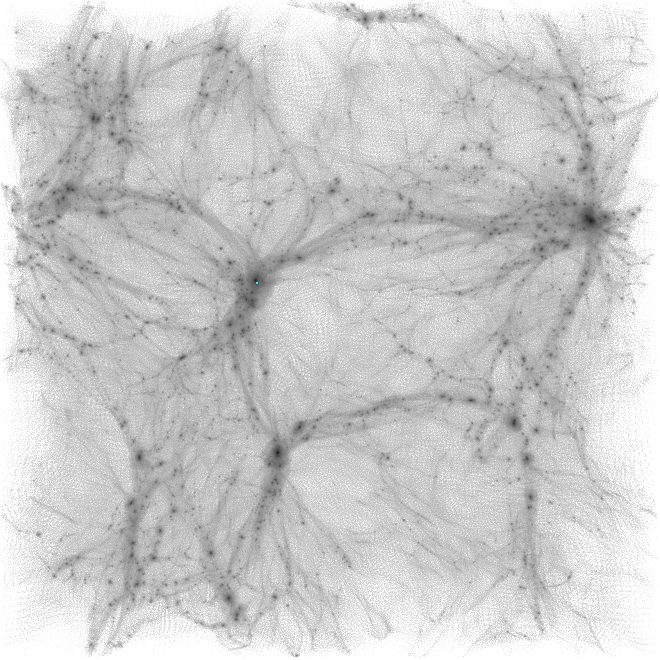}
 \caption{\label{fig:pb15z30pcls} Slice through the simulation box at redshift $z=30$ for the peak-to-background ratio 15 simulation.
 The halo that was seeded by the Gaussian overdensity peak is slightly to the left and to the top of the center, highlighted in light blue
 (color online).}
\end{figure}

Given our choice of $A=0.3$, we can make the typical amplitude of the perturbations 15 times smaller by boosting the standard $\Lambda$CDM power spectrum by a factor of 16 on all scales relevant to our simulation, meaning that the linearised root-mean-square perturbation (excluding the large amplitude central halo) is $4.0$ times larger than it would be in $\Lambda$CDM on all scales.
Starting from such initial conditions we simulate and study the formation and evolution of halos. A snapshot at redshift $z=30$ is
illustrated in Fig.~\ref{fig:pb15z30pcls}.

To determine $\alpha$, the slope of the profile in log-log space, we fit the linear function (\ref{eq:fit_fun}) to the central halo.\footnote{Since halos move around the simulation volume, it is not always obvious which halo formed from the Gaussian overdensity. We 
identified that halo by tracking one of the particles that was initially closest to the center of the simulation volume. At each 
redshift, we looked for the halo that 
contained the tracked particle at its core.}

\begin{figure*}[tb]
\centering
\begin{tabular}{cc}  
\includegraphics[width=0.47\linewidth]{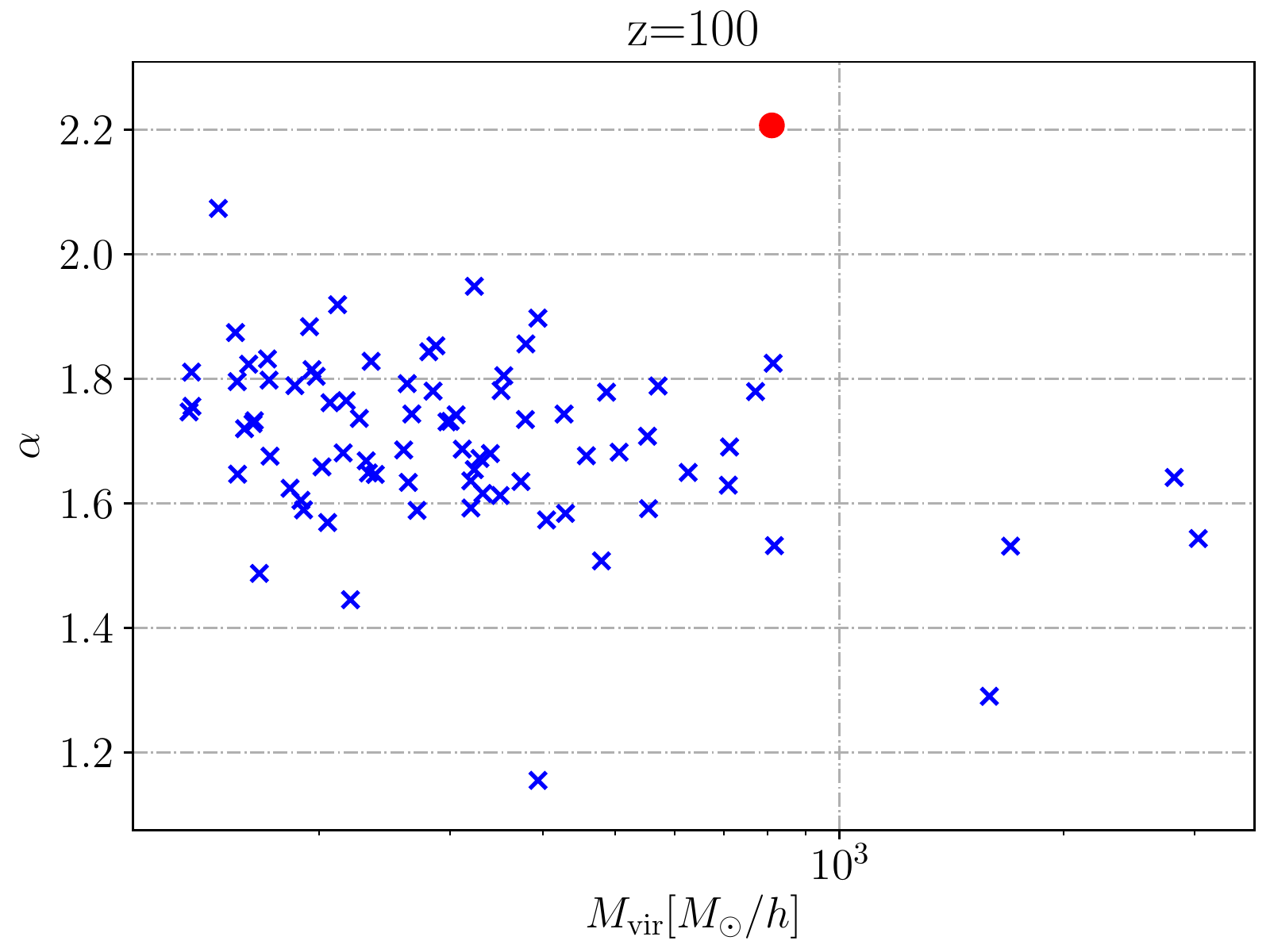}&          
\includegraphics[width=0.47\linewidth]{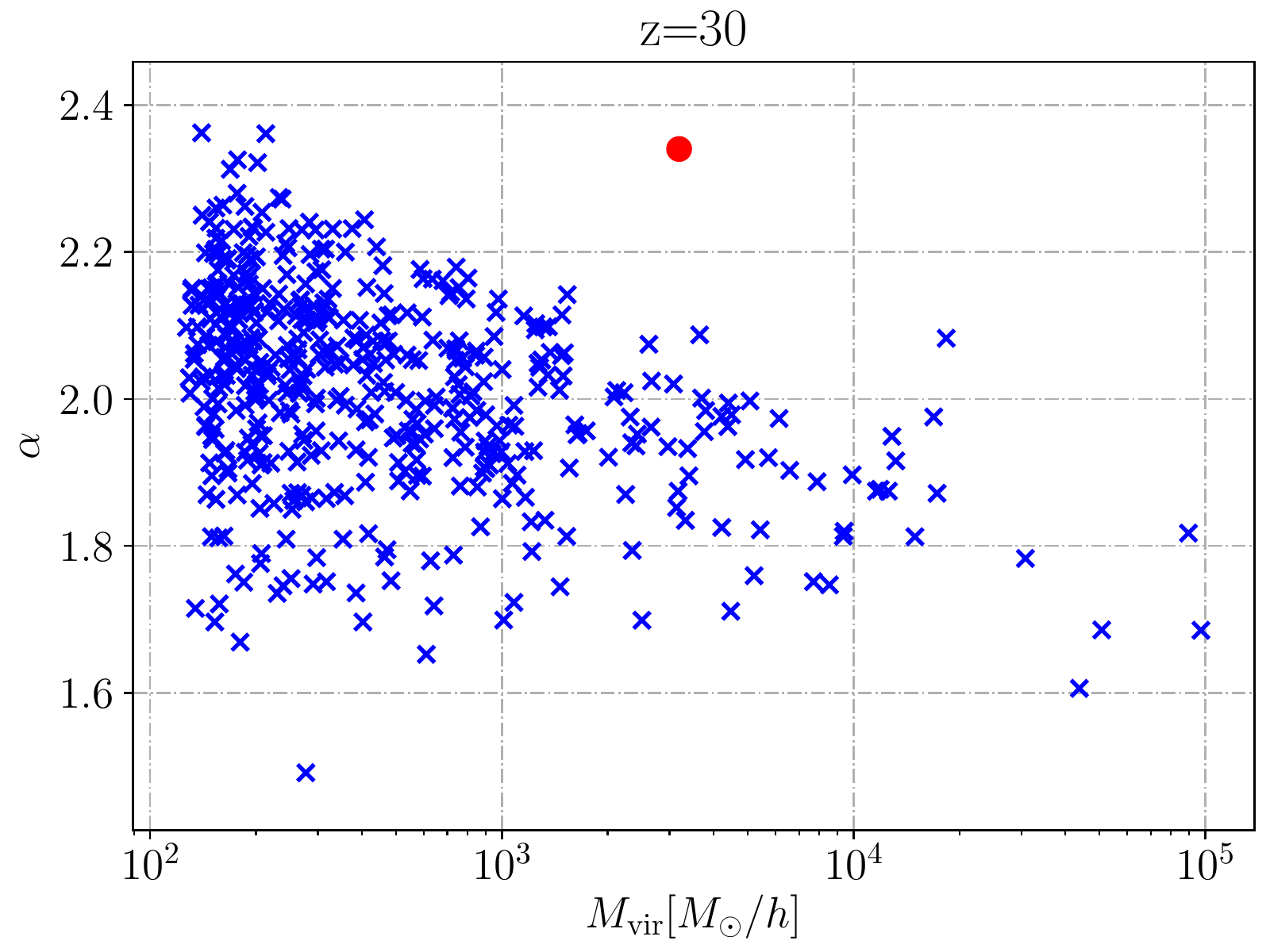}
\end{tabular}
\caption{\label{fig:pbr15alphavsmass}   (Color online) The best-fit power-law exponent $\alpha$ for all halos in the peak-to-background 15 simulation at $z=100$ and $30$. The special halo has an exceptionally steep slope compared to all other halos with a comparable mass.}
\end{figure*}
In Fig. \ref{fig:pbr15alphavsmass} we show a scatter plot of the slope parameter $\alpha$ against the virial mass for every halo in this simulation (at $z=100$ and $z=30$). The red circle in each panel represents the seed halo that started with amplitude $15$ times the background. The seed halo clearly has an exceptional slope compared to its background, even down to $z=30$.

\begin{figure*}[tb]
\centering
\begin{tabular}{cc}  \includegraphics[width=0.47\linewidth]{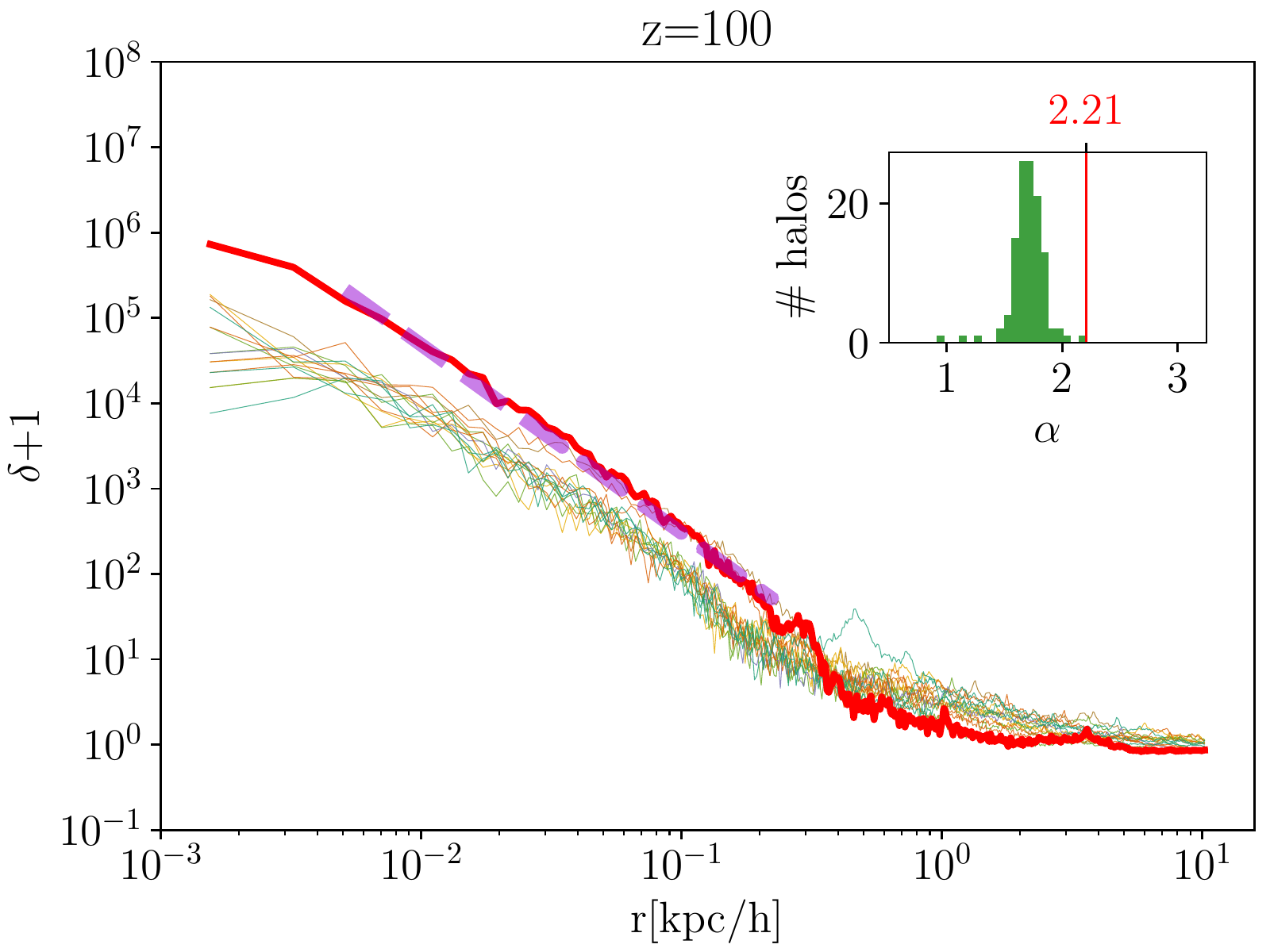}&                                 \includegraphics[width=0.47\linewidth]{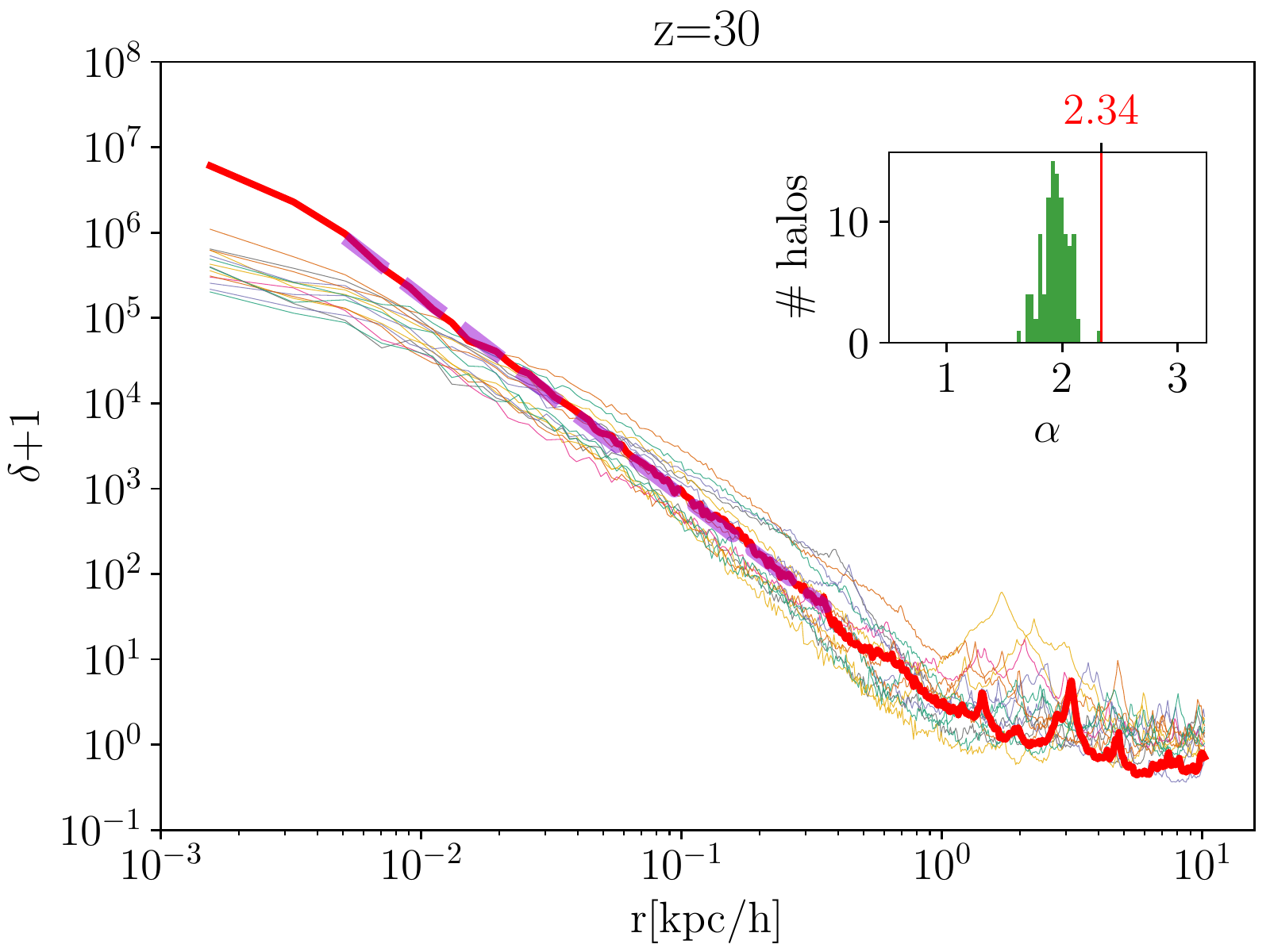}\\[2\tabcolsep]
\includegraphics[width=0.47\linewidth]{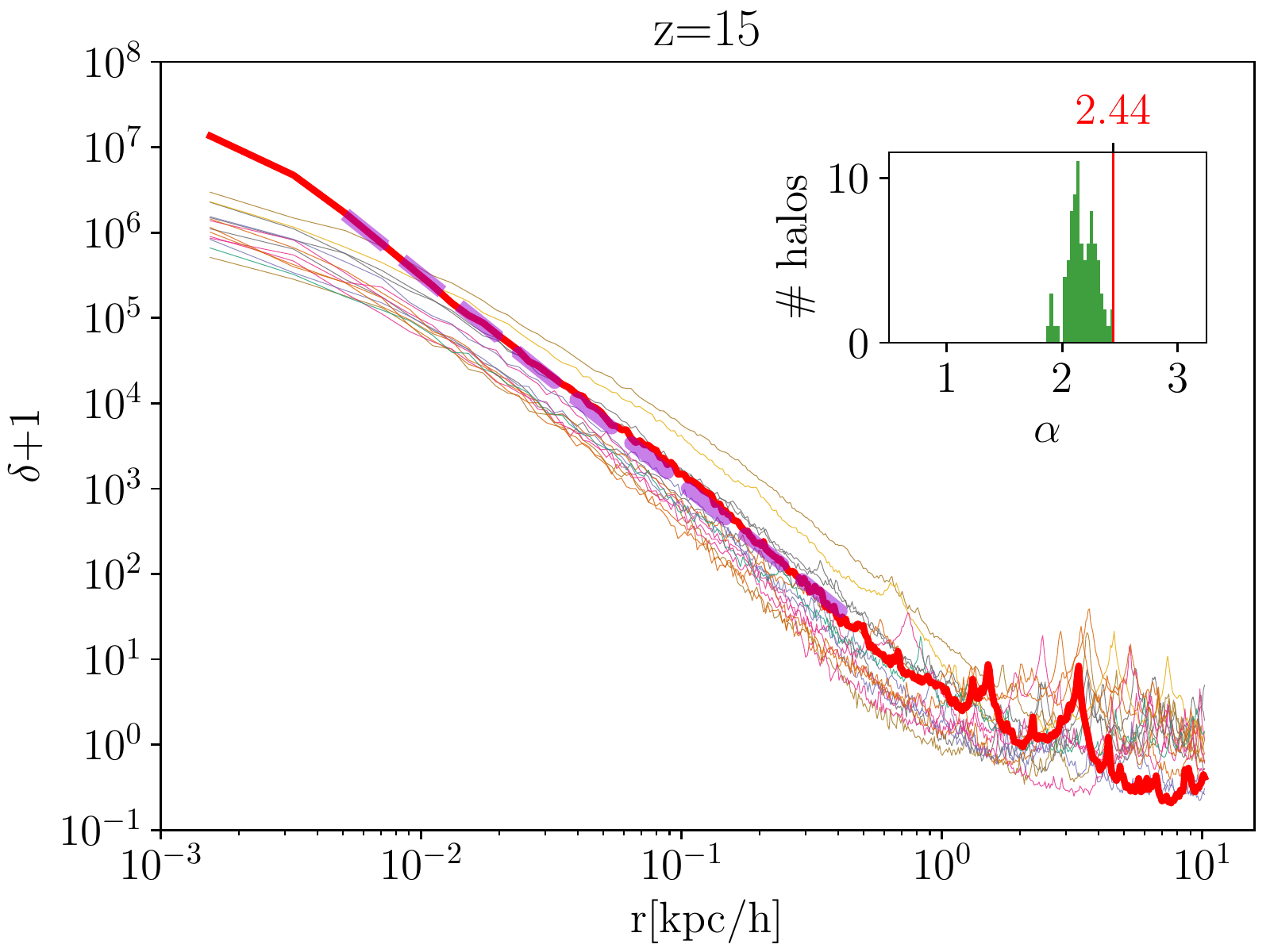}&                                   \includegraphics[width=0.47\linewidth]{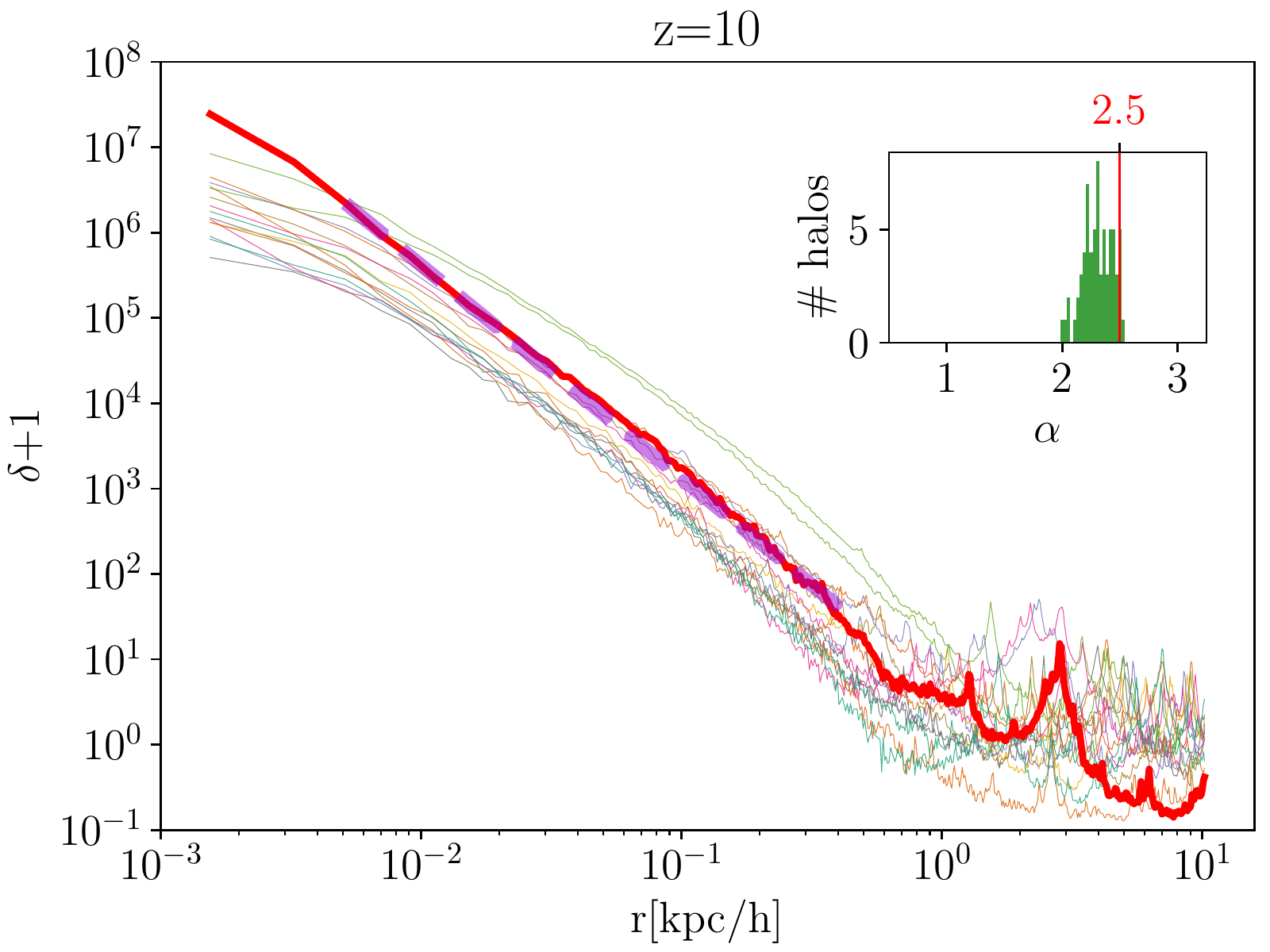}
\end{tabular}
\caption{\label{fig:snr15profiles}   (Color online) The profiles of the halos in the peak-to-background ratio $15$ simulation at $z=100, 30, 15,$ and $10$. The profile of the halo descending from the seed halo is highlighted in red so it can be distinguished from the other halos. For each snapshot, we fit the profiles to eq.~(\ref{eq:fit_fun}) and present the resulting $\alpha$ in a histogram. The range we fit over is $r$ between $0.004\, \kpc/h$ and $r_\vir$ for each halo individually. To plot the profiles, for $z=100$ we only kept halos with $M_\vir > 10^2 M_{\odot}/h$ and for $z=30, 15, 10$ we kept the halos with $M_\vir > 10^3 M_\odot/h$. The power-law fit to the seed halo's profile is plotted as a violet dashed line.}
\end{figure*}
In Fig.~\ref{fig:snr15profiles} we show the density profiles of halos at multiple redshifts. For $z=100$ we show a random sub-sample of
halos with a mass $M_\vir > 10^2 M_\odot/h$ and for the other three redshifts the mass cut is $M_\vir > 10^3 M_\odot/h$. We also include a
histogram of $\alpha$ values for all of the halos that pass this mass cut. It is again clear that amongst the most massive halos the seed
halo's slope remains exceptional, even down to $z=10$ (although less and less so as the background structure develops).

\begin{figure*}[tb]
\centering
\begin{tabular}{cc}
\includegraphics[width=0.47\linewidth]{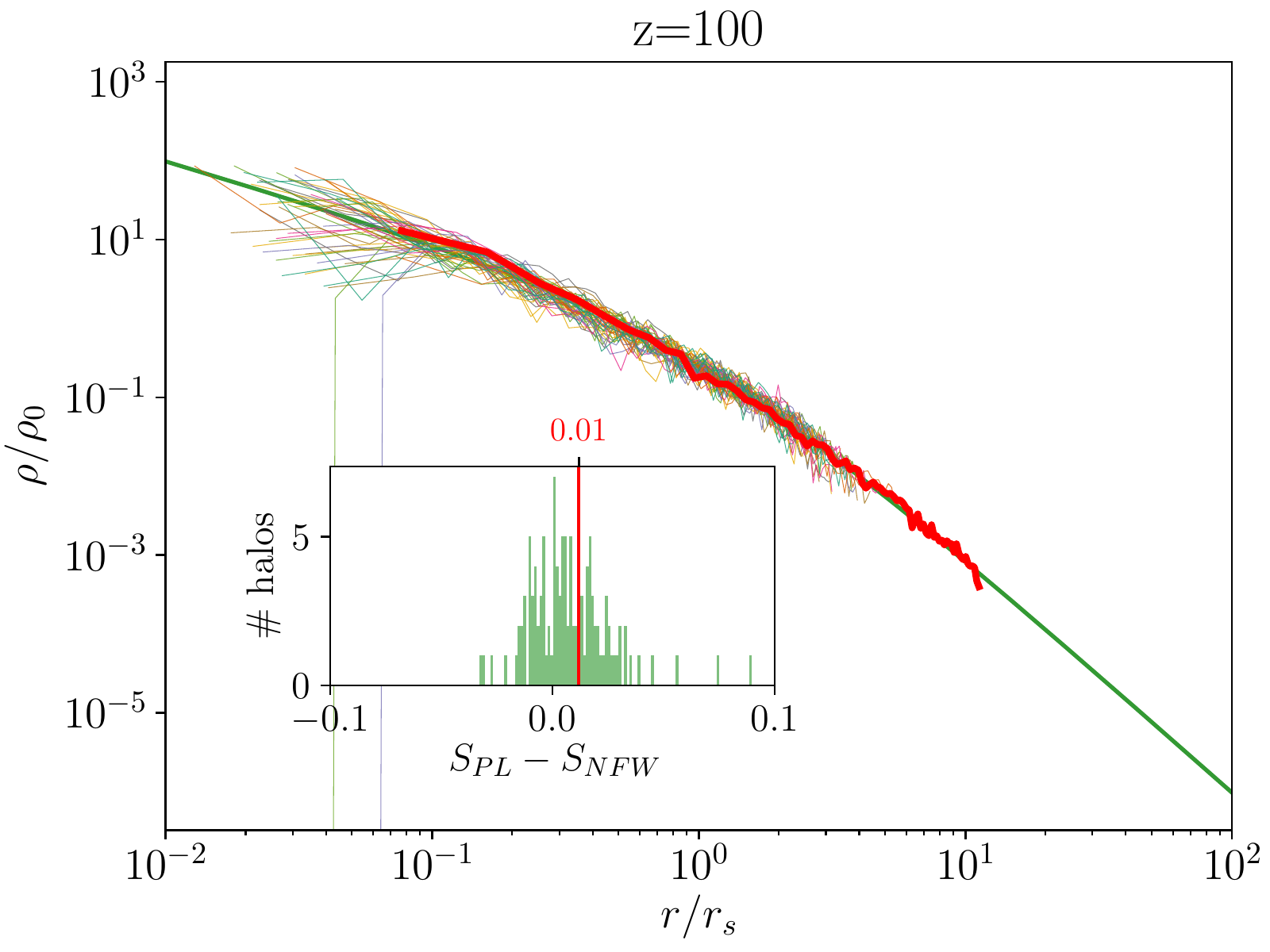}&
\includegraphics[width=0.47\linewidth]{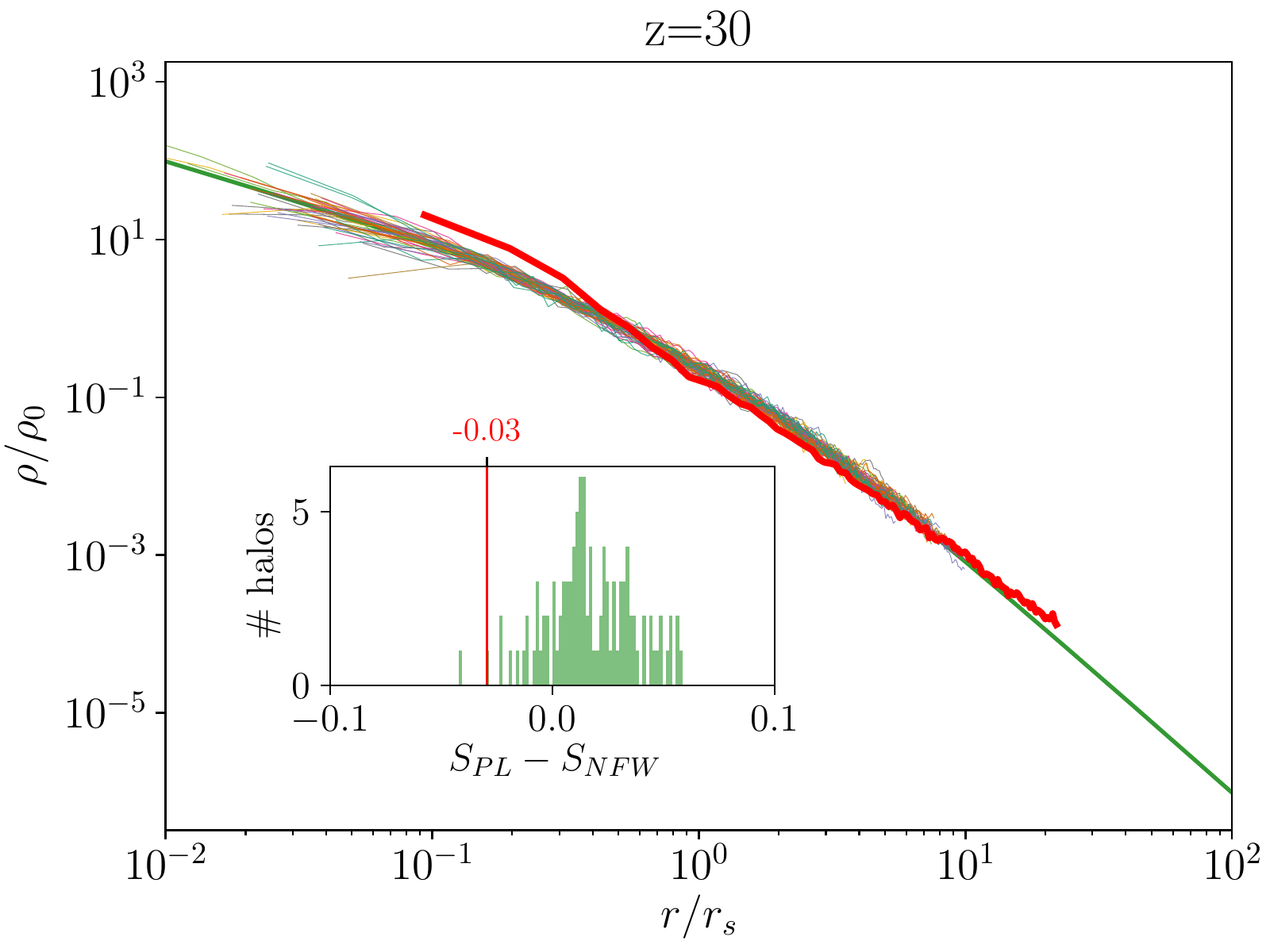}\\[2\tabcolsep]
\includegraphics[width=0.47\linewidth]{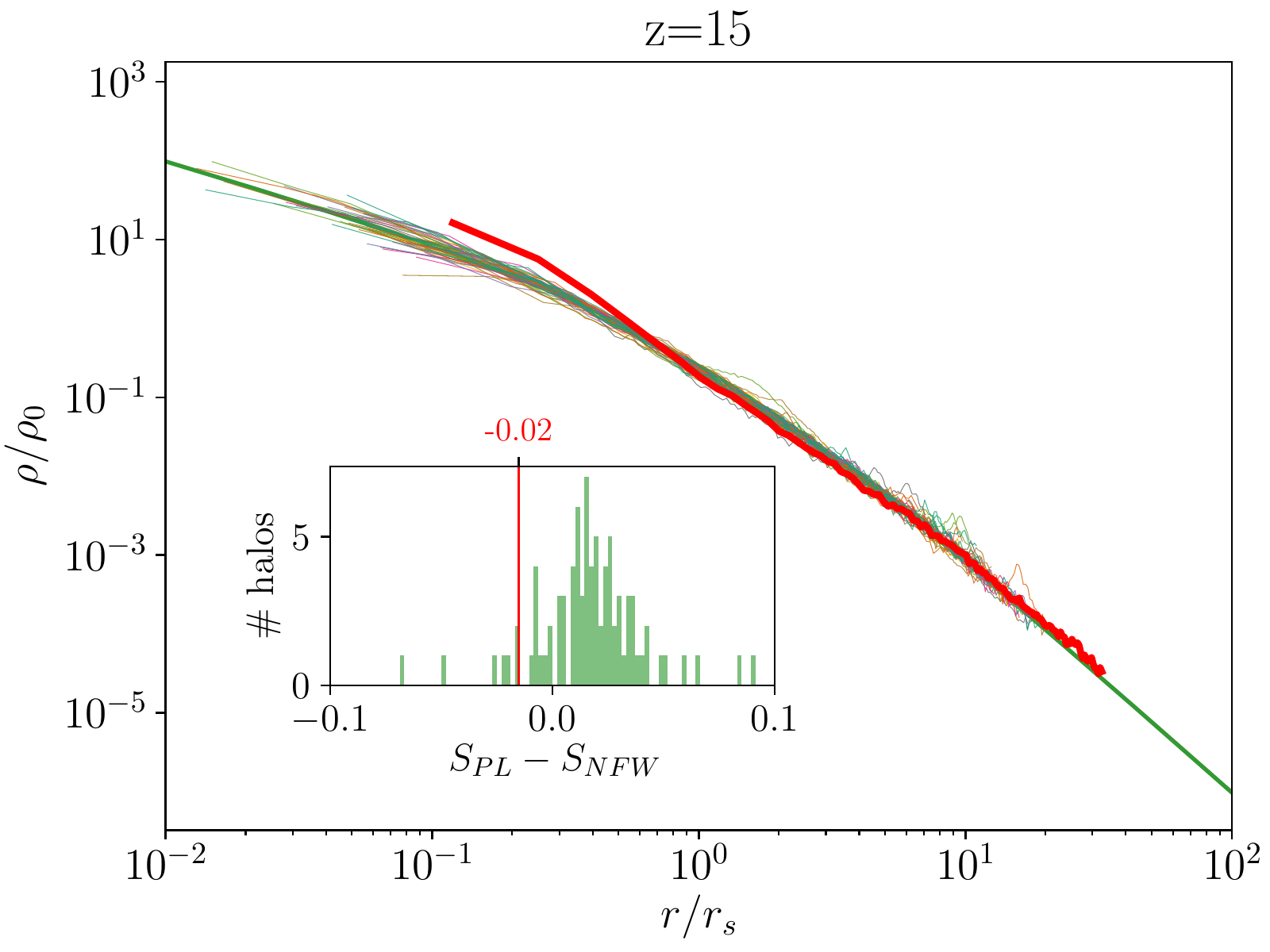}&
 \includegraphics[width=0.47\linewidth]{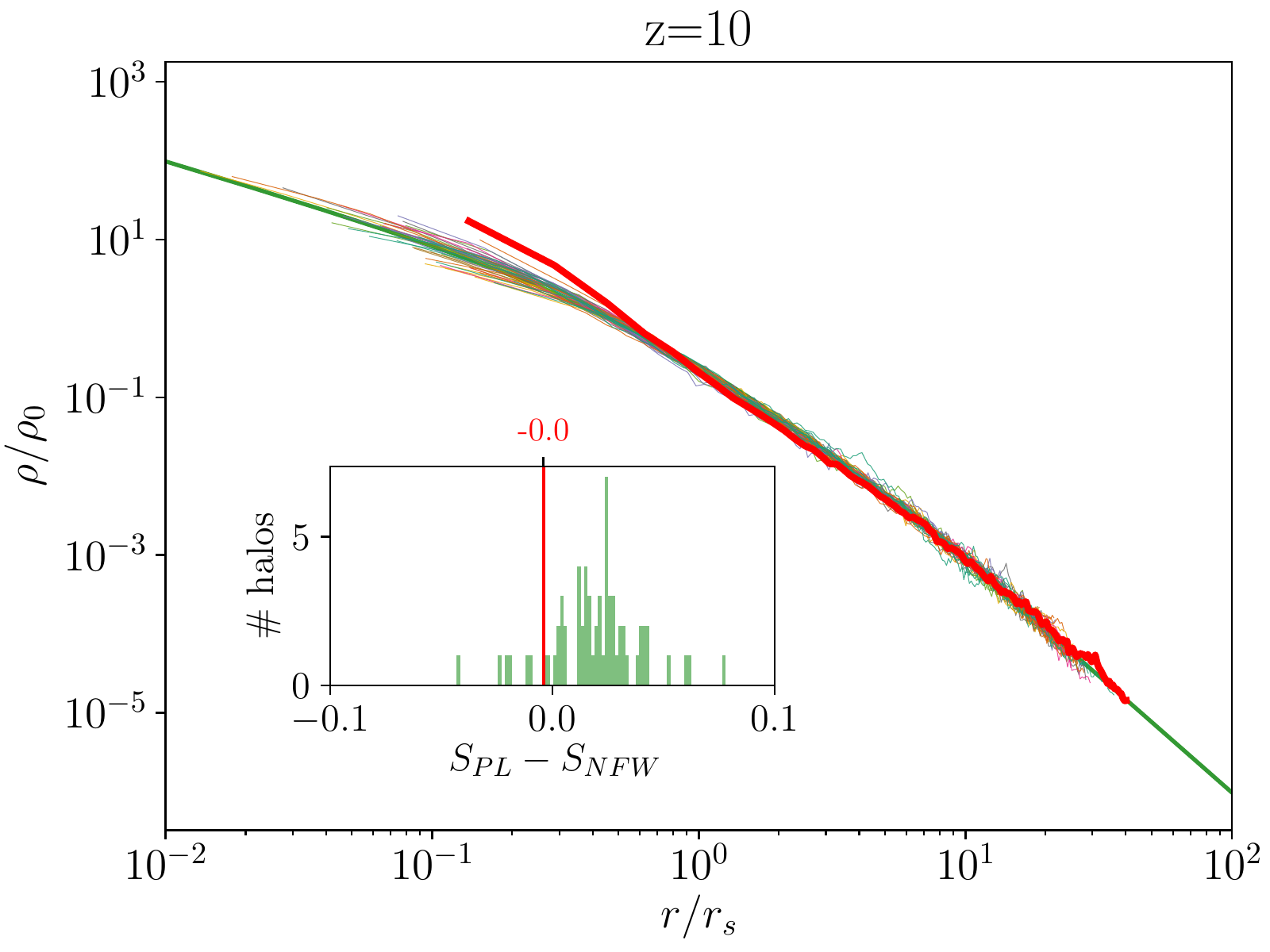}
\end{tabular}
\caption{\label{fig:snr15profilesNFW} (Color online) The rescaled profiles of halos in the peak-to-background ratio $15$ simulation. The NFW analytical prediction is shown in green and the descendant of the seed halo is highlighted in red. Histograms show the difference of a measure for goodness of fit between the power-law and NFW profile. A negative value of this quantity points towards the power-law being a better fit than NFW. With the exception of $z=100$, the special halo seems to favour power-law profile over NFW. The exceptionality is particularly noticeable near the center of the halo. Fitting the special halo with an NFW profile gives a very small $r_{\rm s}$. Here this is manifested by the special halo's profile being shifted towards larger radii than any other halo.}
\end{figure*}
Finally, in Fig.~\ref{fig:snr15profilesNFW}, we collapse the halo profiles on to the NFW function in the manner described at the end of section \ref{sec:fitting} (we apply the same mass cuts as for Fig.~\ref{fig:snr15profiles}). We also show a histogram of the difference in the value of our fitting statistic $S$, eq.~\eqref{eq:fittingstat} for the best fitting power-law and NFW profiles (i.e. $S_{\rm PL} - S_{\rm NFW}$). It is clear from the profiles that the NFW profile is not an excellent fit for the special seed halo. There is a noticeable upturn at small $r/r_s$. Also note how far to the right the special halo's profile extends in each panel. This is a consequence of the NFW fit requiring a small $r_s$ value, exceptionally smaller than any other halo that satisfies the mass cuts.

We cannot resolve the profile of any halo on scales below the resolution limit of our simulation. However, because the special halo remains exceptional compared to its background, even down to $z=10$, it seems reasonable to expect that the profile would remain close to a power-law even beyond the limits of our resolution.

\subsection{Peak-to-background ratio 5}\label{sec:pb5}\label{sec:sn5}

\begin{figure}
 \includegraphics[width=\columnwidth]{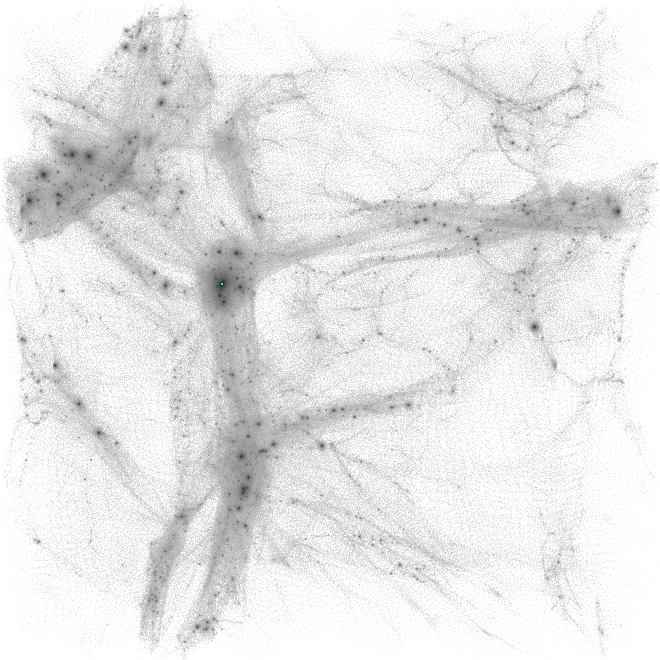}
 \caption{\label{fig:pb5z30pcls} Slice through the simulation box at redshift $z=30$ for the peak-to-background ratio 5 simulation.
 The halo that was seeded by the Gaussian overdensity peak is slightly to the left and to the top of the center, highlighted in light blue
 (color online).}
\end{figure}
We now increase the amplitude of the surrounding perturbations even more and repeat the analysis. In the results of this subsection the central overdensity is only $5$-times larger than a typical one. A snapshot at redshift $z=30$ is shown in Fig.~\ref{fig:pb5z30pcls}.

\begin{figure*}[tb]
\centering
\begin{tabular}{cc}  \includegraphics[width=0.47\linewidth]{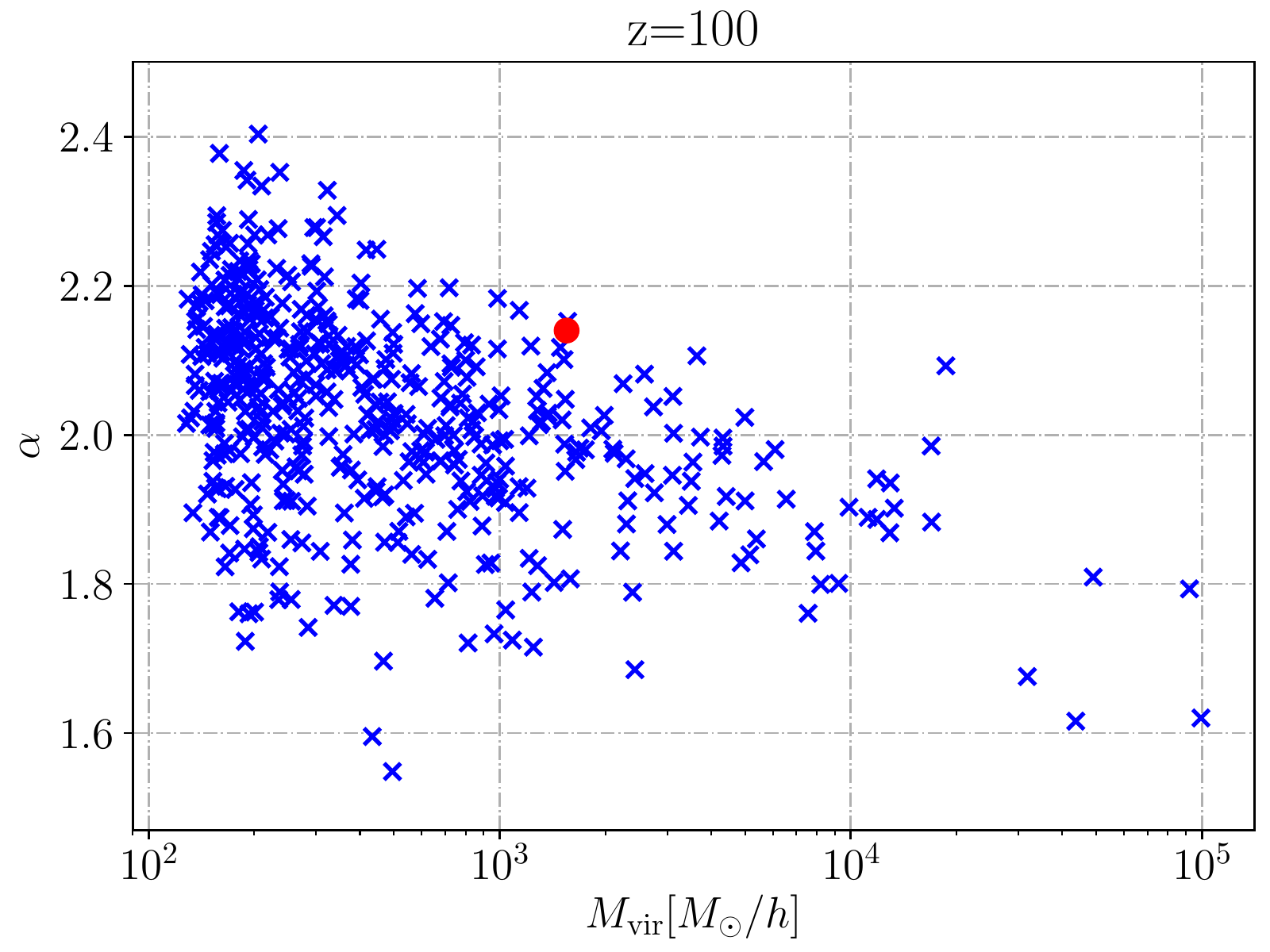}&            \includegraphics[width=0.47\linewidth]{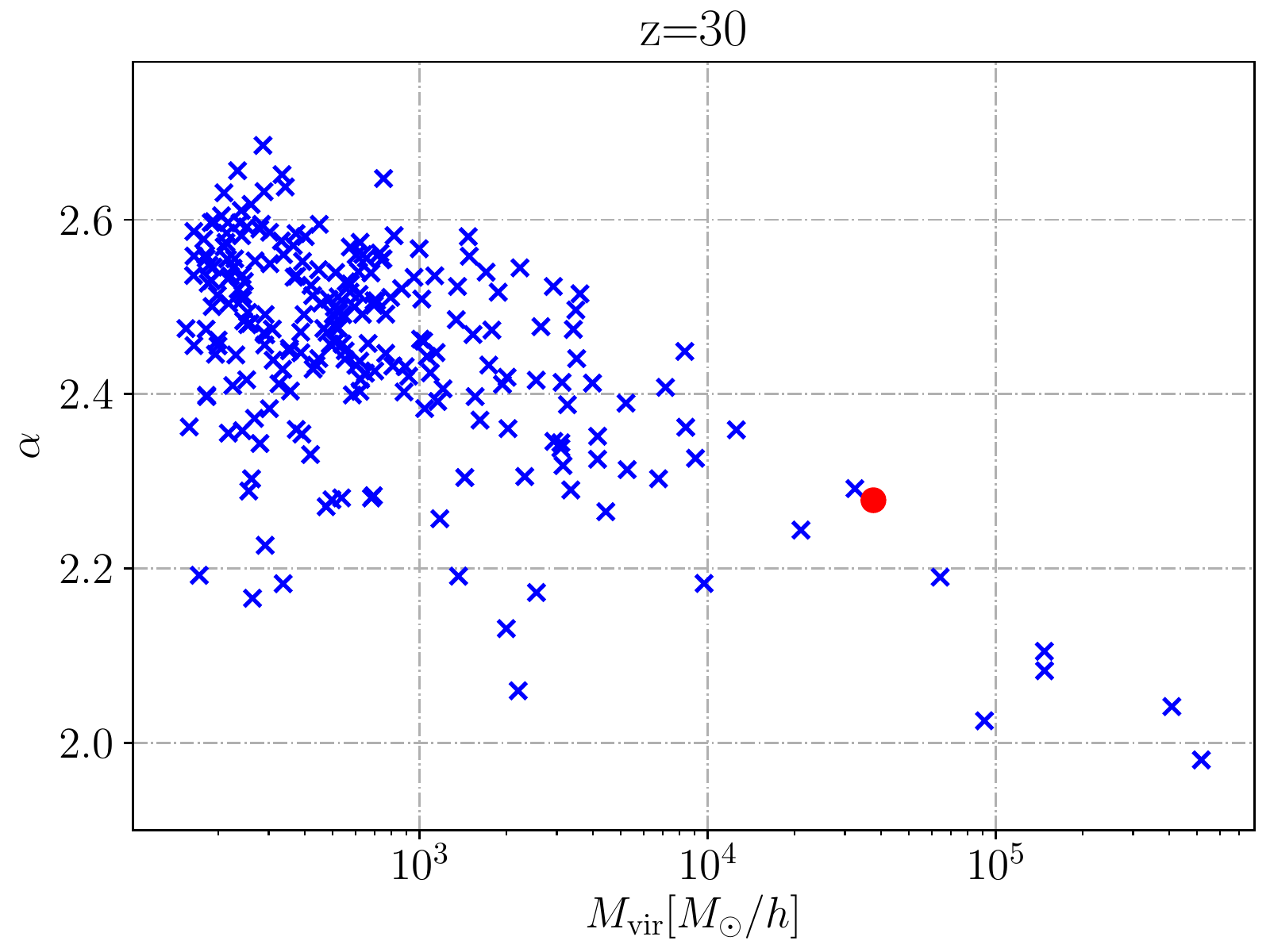}
\end{tabular}
\caption{\label{fig:pbr5alphavsmass}   (Color online)  Figures analogous to Fig.~\ref{fig:pbr15alphavsmass}, but for the peak-to-background-ratio 5. The power-law parameter $\alpha$ of the halo, formed from the special seed, is no longer very different from the other halos.}
\end{figure*}
\begin{figure*}[tb]
\centering
\begin{tabular}{cc}
\includegraphics[width=0.47\linewidth]{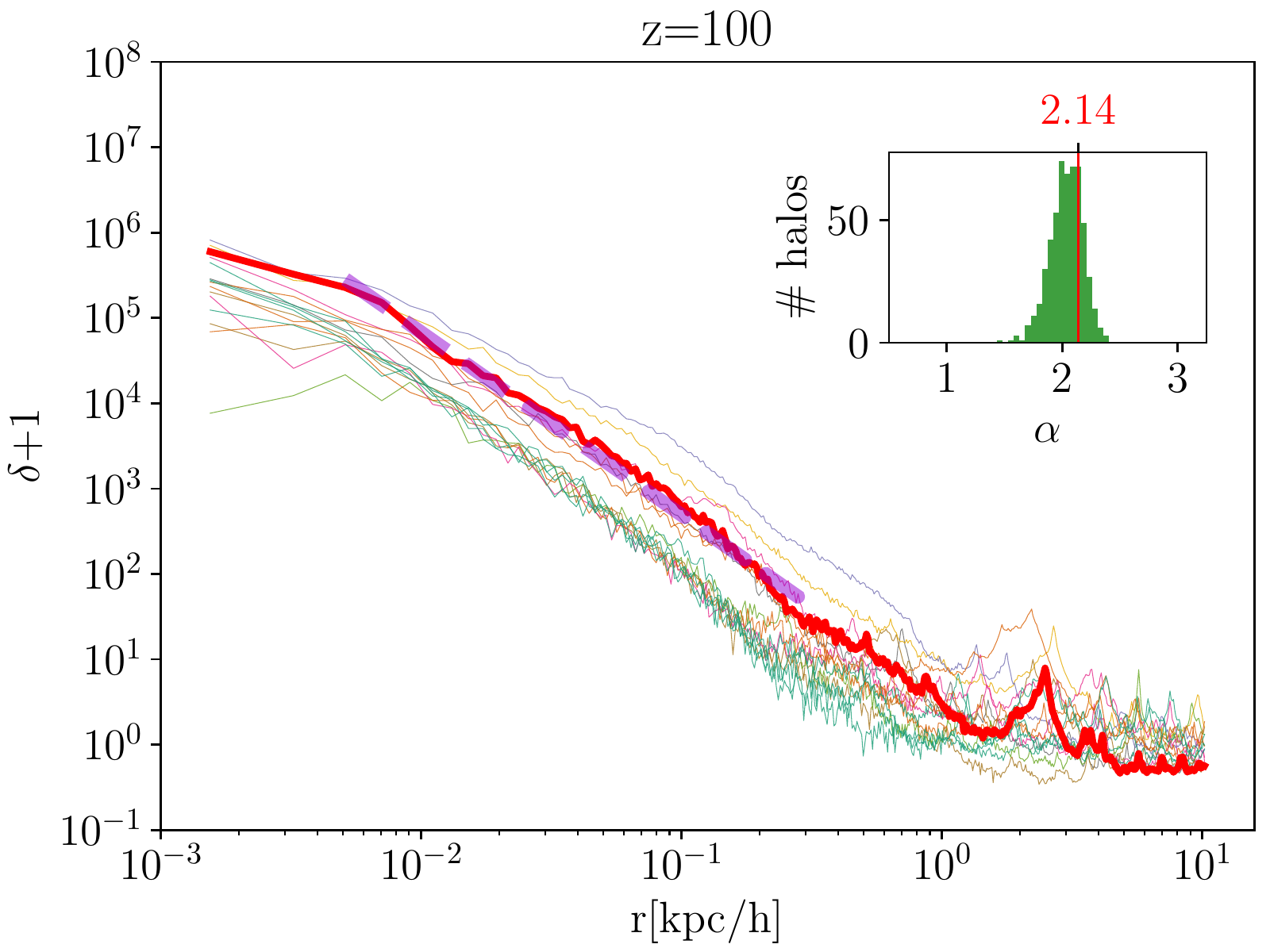}&
\includegraphics[width=0.47\linewidth]{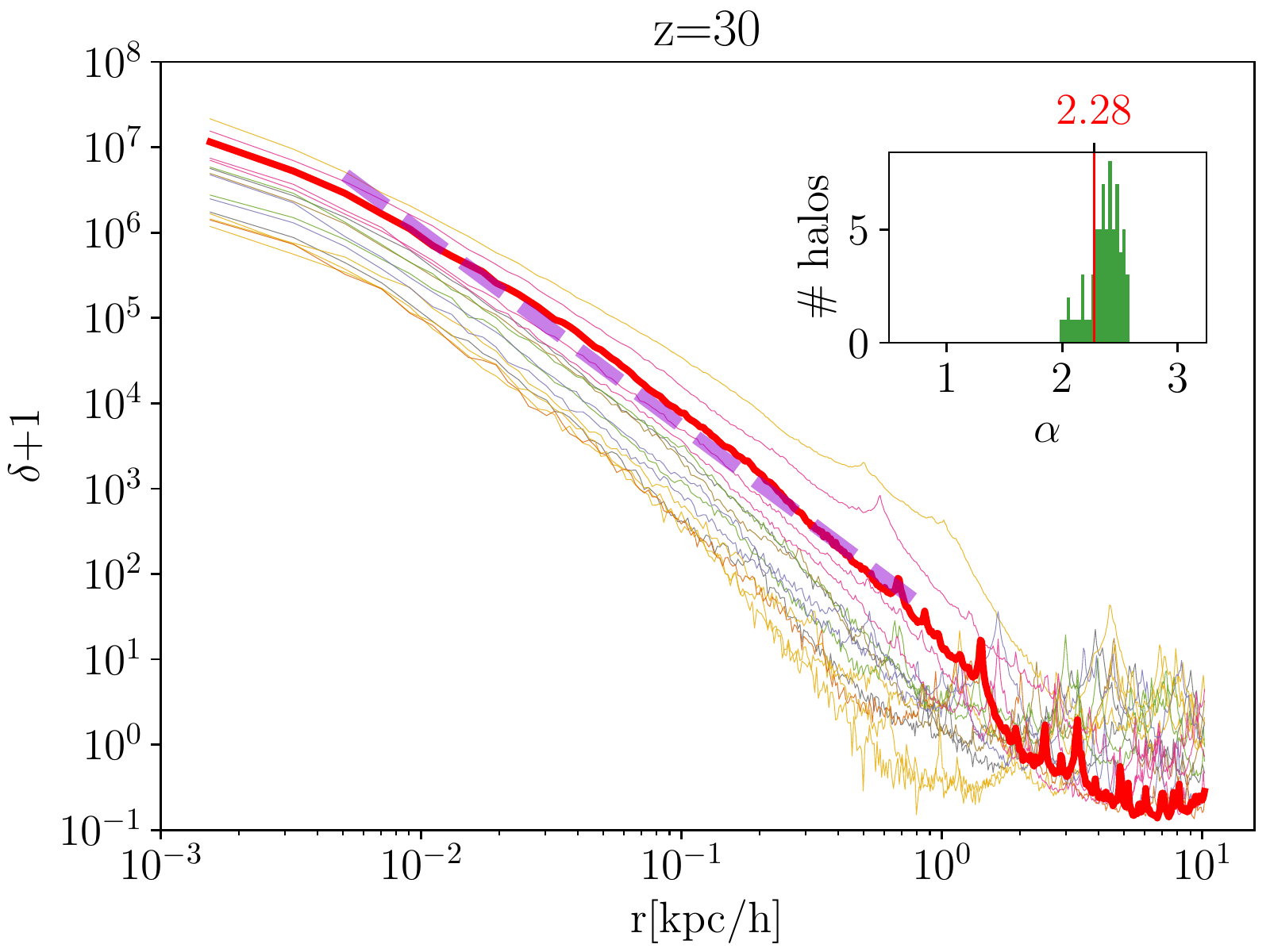}
\end{tabular}
\caption{\label{fig:snr5profiles} (Color online) The profiles of the halos in the peak-to-background ratio $5$ simulation at $z=100$ 
and $30$. We have again only included halos with $M_\vir>10^2 M_\odot/h$ for the $z=100$ profiles and $M_\vir>10^3 M_\odot/h$ for $z=30$. 
The profile of the descendant of the seed halo is again highlighted in red. Unlike the situation in the peak-to-background 15 simulation, the slope of the special halo is no longer more extreme than the other halos in the box.}
\end{figure*}
\begin{figure*}[tb]
\centering
\begin{tabular}{cc}
\includegraphics[width=0.47\linewidth]{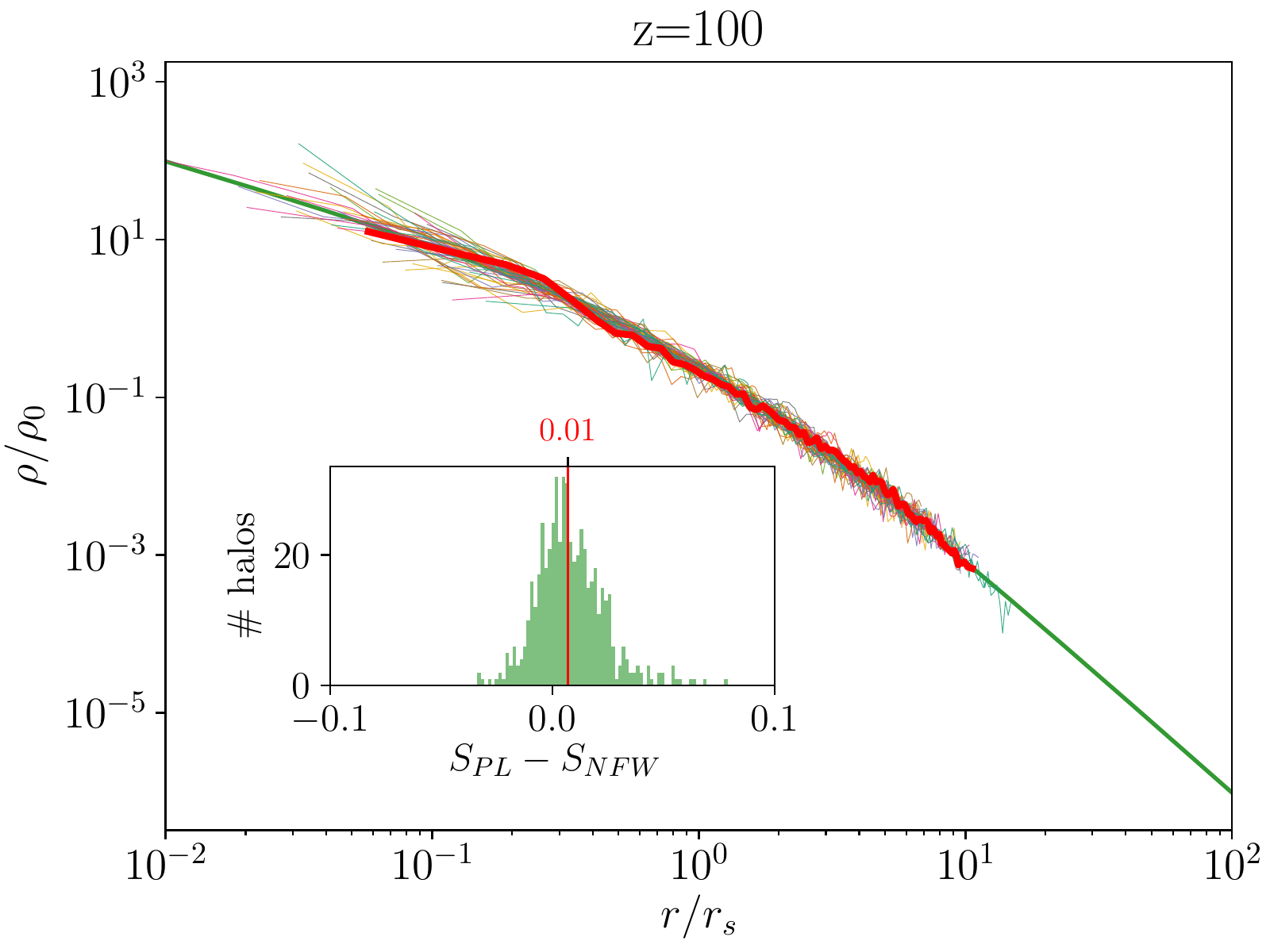}&
\includegraphics[width=0.47\linewidth]{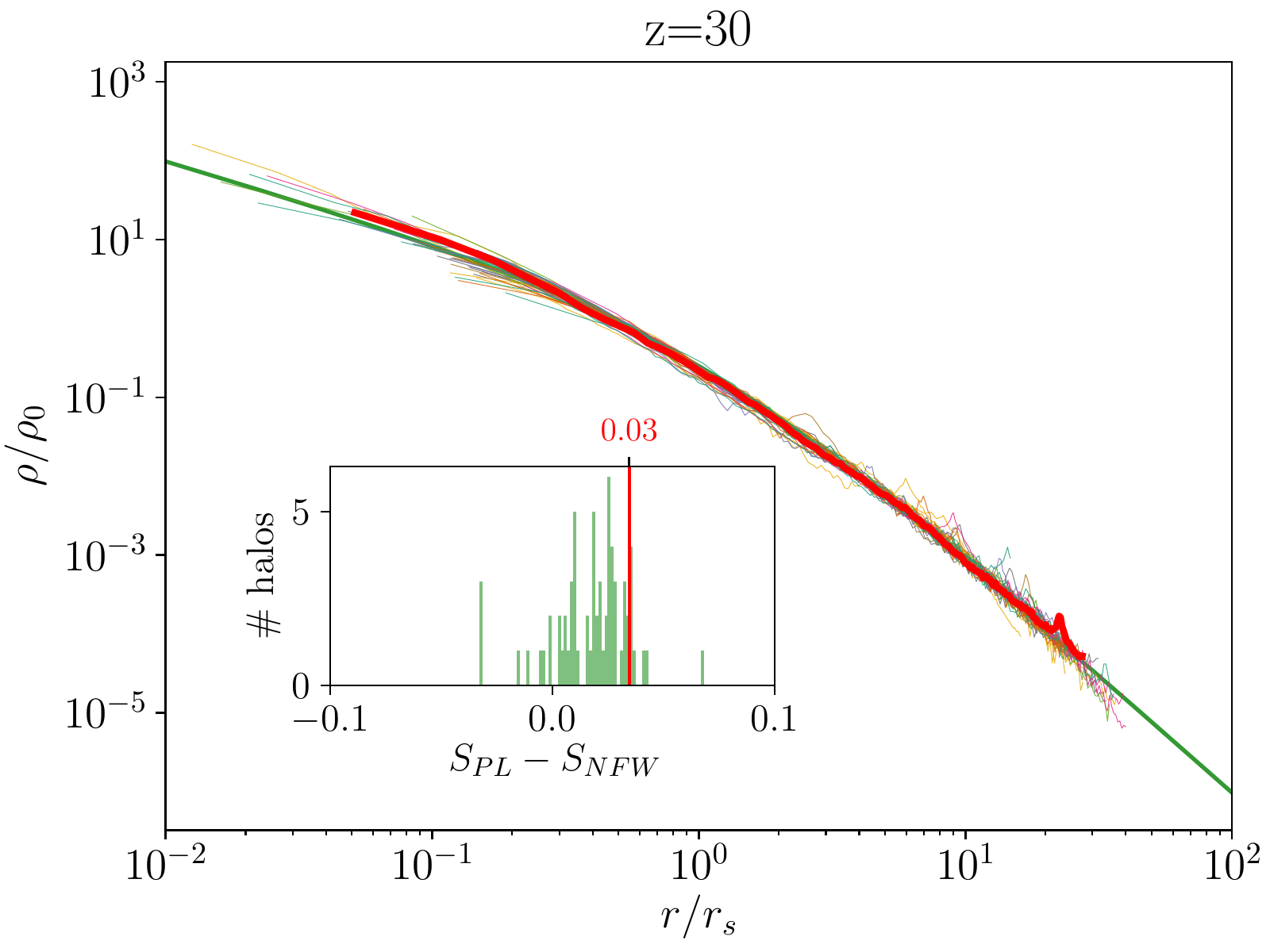}\\
\end{tabular}
\caption{\label{fig:snr5profilesNFW} (Color online) The rescaled profiles of halos in the peak-to-background ratio $5$ simulation. The analytical prediction is shown in green and the special halo is highlighted in red. In contrast to the peak-to-background ratio $15$, the special halo does not appear distinguishable from the rest. The $S_{\rm PL} - S_{\rm NFW}$ measure also demonstrates the profile of the special halo to be a better fit to NFW than to power-law (note a smaller value of $S$ indicates a better fit, see equation \eqref{eq:fittingstat} and the discussion around it).
  }
\end{figure*}
Figs.~\ref{fig:pbr5alphavsmass}, \ref{fig:snr5profiles} and \ref{fig:snr5profilesNFW} are analogous to Figs.~\ref{fig:pbr15alphavsmass}, \ref{fig:snr15profiles} and \ref{fig:snr15profilesNFW}, however for the simulations with  even larger background perturbations and showing only $z=100$ and $z=30$.

The most striking observation to make is that the special halo is no longer at all exceptional compared to its surroundings. Its slope compared to its mass is large compared to other halos in the simulation, but not exceptionally so. Moreover, its value of $S_{\rm PL} - S_{\rm NFW}$, although still favouring a power-law, is not at all special compared to other halos, some of which also favour the power-law.

We cannot say what happens below the limits of our resolution; however it is clear that when an initial, spherically-symmetric perturbation is embedded in a background $5$ times smaller it is no longer exceptional at least by $z=100$. Therefore it does not behave as if it was in a homogeneous background and it is reasonable to assume that it has not 
formed a UCMH-like profile on smaller scales. Although we do not show it here, this holds even at $z=300$.

  \subsection{Summary of special halo simulations}
In this section we have shown that as the background of an initial spherically-symmetric perturbation is increased, the halo descending from this seed perturbation becomes less and less exceptional. This might not appear too surprising, however it isn't trivial that an initial $5 $-$\sigma$ fluctuation at $z=10\,000$ will be entirely unexceptional by $z=100$.

  The conclusion we take from this section is that  unless a fluctuation has an initial amplitude at least between 5 and 15 times larger than the typical background  fluctuations its halo will soon become comparable in slope, mass and density to many other nearby halos. Unless the perturbation is more extreme initially than one would expect to find given Gaussian initial conditions, see Sec.~\ref{sec:extreme},  it will not grow into a UCMH-like halo.
  
We stress that this is not the result of lowering the density of the initial seed fluctuation. The initial profile was identical in all three subsections; the only thing that was changed was the amplitude of background fluctuations. Therefore, when examining the evolution of objects on the very small scales in the early Universe, a single density contrast $\delta_c$ is not enough to describe the subsequent evolution. The evolution depends sensitively on the environment and therefore any constraints must too.

\section{Boosting the power spectrum around the $1\,\kpc/h$ scale}
\label{sec:PSboost}

In this section we no longer consider the evolution of a specific overdensity peak that was planted \textit{by hand}. Instead,
we increase the probability of extreme \textit{random} fluctuations on a similar scale by boosting the variance of the primordial
perturbation modes over some range of scales.
  In principle this boost can take many forms (e.g. a step, or a bump in Fourier space) and the precise nature of structure formation will depend on the form 
  of the boost. However, UCMHs are claimed to form with a sufficiently large abundance to be observed under \emph{any} boost of the power 
  spectrum with sufficiently large amplitude.
  
  We see no halo with convincing UCMH properties; however we are somewhat limited by the resolution. We do however see many compact structures. 
  With hindsight, this is not surprising either. If we boost the primordial spectrum then structures on the boosted scales form earlier. 
  This means these structures form when the Universe is denser. Therefore they also reach a larger virial density than they would have 
  had they formed from a non-boosted initial power spectrum.

  Note however this is not only true for the most extreme, and therefore rare, 
  structures forming under a boosted power spectrum. In fact \emph{every} structure that forms on the boosted scales forms earlier and is 
  therefore more compact.
  As a consequence, even though we do not find structures as compact as a hypothetical UCMH, 
  the structures we find may, due to their increased abundance, have their own unique cosmological signals.
  It would be useful and interesting future work to examine how the new type of compact halos we describe depend on the specific nature of the boost to the power spectrum.
  
  The form of our boost in the power spectrum is as follows. Firstly we take the following unboosted power-law power spectrum of the primordial curvature perturbation $\zeta$, $\mathcal{P}^\zeta_0(k)=A_s \left(k/k_\mathrm{pivot}\right)^{n_s-1}$ with $A_s=2.26\times 10^{-9}$, $k_\mathrm{pivot}=0.05\, \, \mathrm{Mpc}^{-1}$ and $n_s = 0.96$.

We then boost this power spectrum to form the primordial power spectrum used for our simulations, $\mathcal{P}^\zeta(k)$. Specifically,
\begin{equation}\label{eq:boostedPk}
\mathcal{P}^\zeta(k)=\mathcal{P}^\zeta_0(k)\left(1+B(k)\right)
\end{equation}
where
\begin{equation}\label{eq:boost}
B(k)=A_b\exp\left(- 2.77 \left(\ln{k}-\ln{k_\star}\right)^2 \right).
\end{equation}
The scale at which we boost the simulation's power spectrum is $k_{\star}=1 \,\,h \mathrm{kpc}^{-1}$. 

The value of $2.77$ in the definition of $B(k)$ is chosen such that the full width at half maximum of the boost is 1 (in units of $\ln{k}$). This means that our power spectrum is boosted over a width corresponding to approximately one efolding, which is a natural length scale during inflation.

\begin{figure}
\includegraphics[width=\columnwidth]{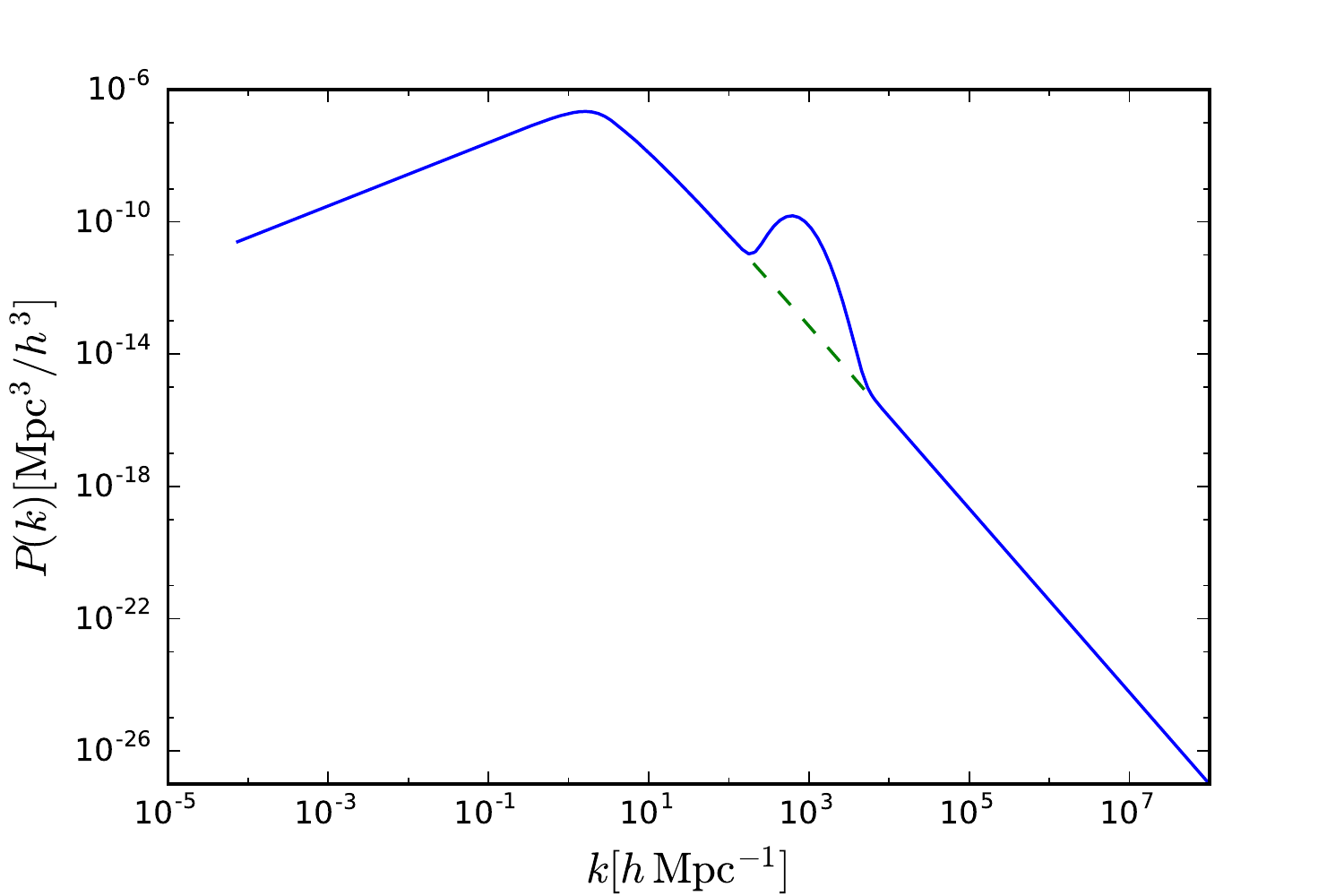}
\caption{\label{fig:boostedpk} Boosted matter power spectrum used in our simulations. Plot shown has boost parameters $A_b=10^3$ and $k_\star=1\, \,  h \mathrm{kpc}^{-1}$ from eq.~\eqref{eq:boost}. The matter power spectrum has been calculated using CAMB and is outputted at $z=100\,000$. Note that this is only the dark matter power spectrum.}
\end{figure}

Before starting each simulation we use \textit{CAMB} \cite{Lewis:1999bs} to calculate the linear transfer functions and hence the matter power spectrum at the simulation's starting redshift. The shape of such a boosted power spectrum is plotted in Fig.~\ref{fig:boostedpk} for $A_b=10^3$, $k_\star= 1 \, \, h \mathrm{kpc}^{-1}$  and at redshift $z=100\,000$. Note that this is the power spectrum of dark matter only. The power spectrum of baryons is significantly different due to baryons still being tightly coupled to photons at this high redshift.

\begin{figure}
\includegraphics[width=\columnwidth]{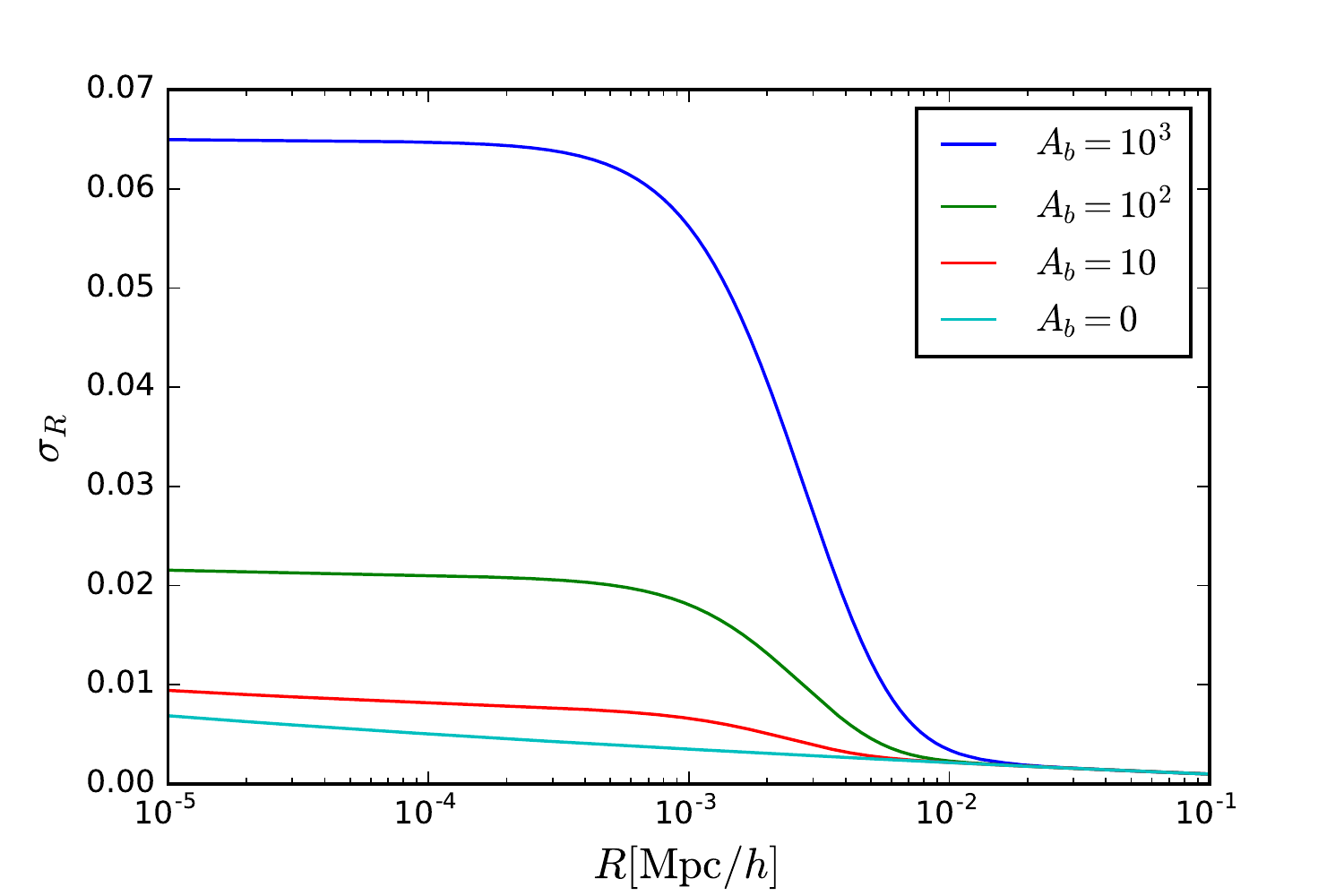}
\caption{\label{fig:boostedsigR} The root-mean-square fluctuations of the density perturbation when smoothed with a tophat filter of radius $R$ (i.e $\sigma_R$). This is for the same input power spectrum as that shown in Fig.~\ref{fig:boostedpk} with four different boosts and at the same redshift.}
\end{figure}

We also show $\sigma_R$ (i.e. the rms of the density contrast smoothed by a spherical tophat on a scale $R$) in Fig.~\ref{fig:boostedsigR} for the same input power spectrum with four different boost amplitudes at the same redshift.

We ran simulations with boosts of $A_b=0,10,100$ and $1000$. Each simulation was given the same seed. At $z=100\,000$ the largest fluctuation in the $A_b=1000$ simulation, when the density was smoothed with a spherical tophat filter of radius $1\, \mathrm{kpc}/h$ was $\delta_R=0.23$. With this boost, $\sigma_R$ at this radius and redshift is $0.054$ therefore this corresponds to a $4.3 $-$\sigma$ fluctuation.

\begin{figure*}[tb]
\centering
\begin{tabular}{cc}
    \includegraphics[width=0.47\linewidth]{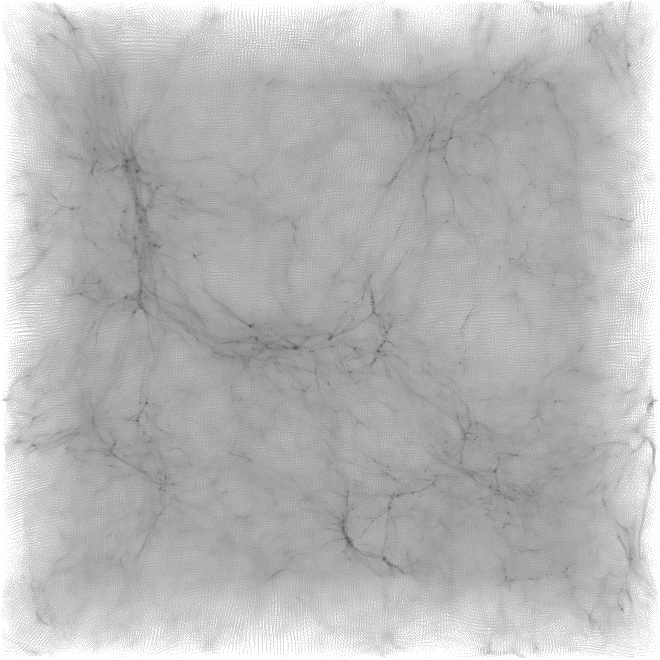}&
    \includegraphics[width=0.47\linewidth]{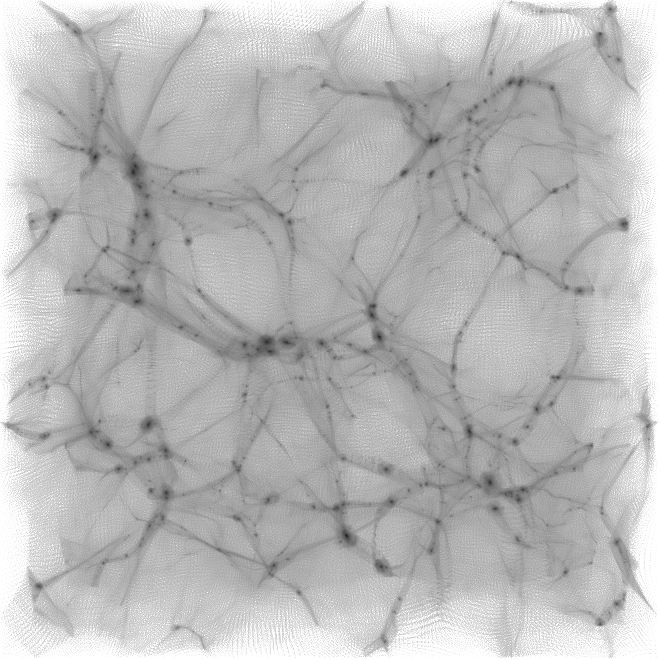}\\[2\tabcolsep]
    \includegraphics[width=0.47\linewidth]{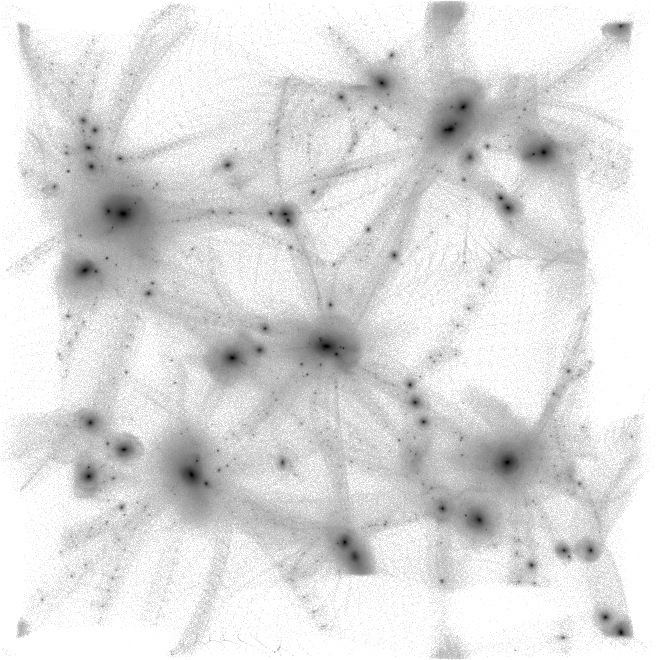}&
    \includegraphics[width=0.47\linewidth]{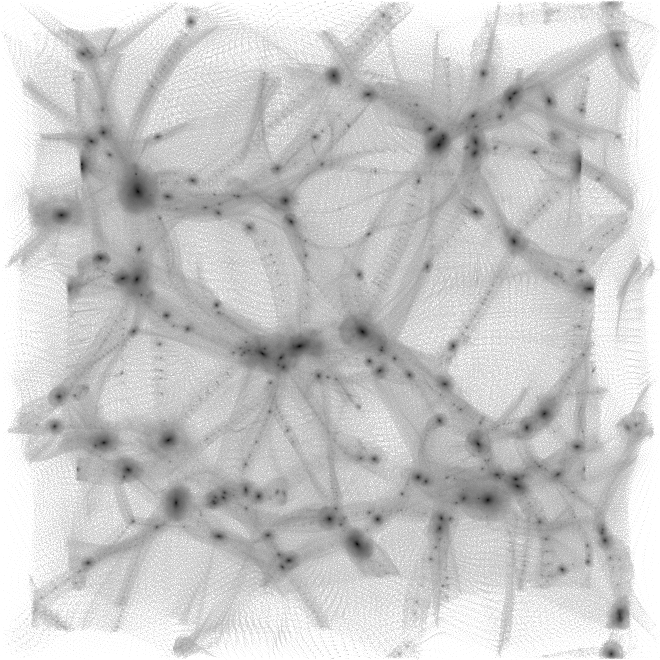}\\[2\tabcolsep]
\end{tabular}
\caption{\label{fig:boostz30pcls} Slices through the simulation volume at redshift z = 30 for the simulations with boosted power spectrum. The amplitudes of the boosts are $A_b = 0, 10, 100, 1000$, clockwise, starting from the top left.}
\end{figure*}

We start each simulation at $z=5 \times 10^6$ and run them to $z=15$. We do not go to lower redshifts than this because at $z=10$ the scale of our box, $32 \,\kpc/h$, becomes non-linear. Snapshots at redshift $z=30$ are illustrated in Fig.~\ref{fig:boostz30pcls}.

\subsection{Results of boosted simulations}\label{sec:NFW-fit}
\begin{figure*}[tb]
\centering
\begin{tabular}{ccc}
    \includegraphics[width=0.31\linewidth]{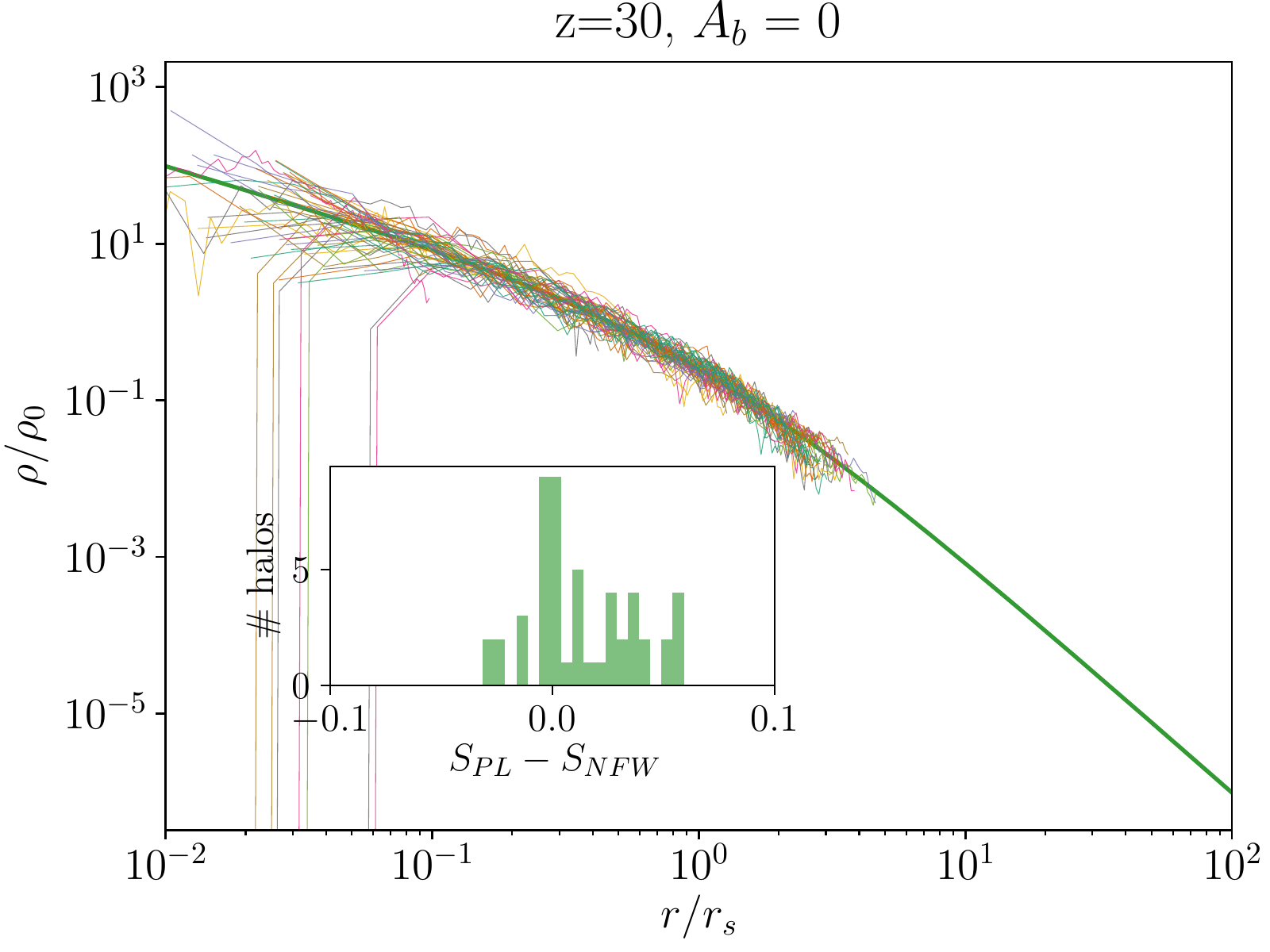}&
    \includegraphics[width=0.31\linewidth]{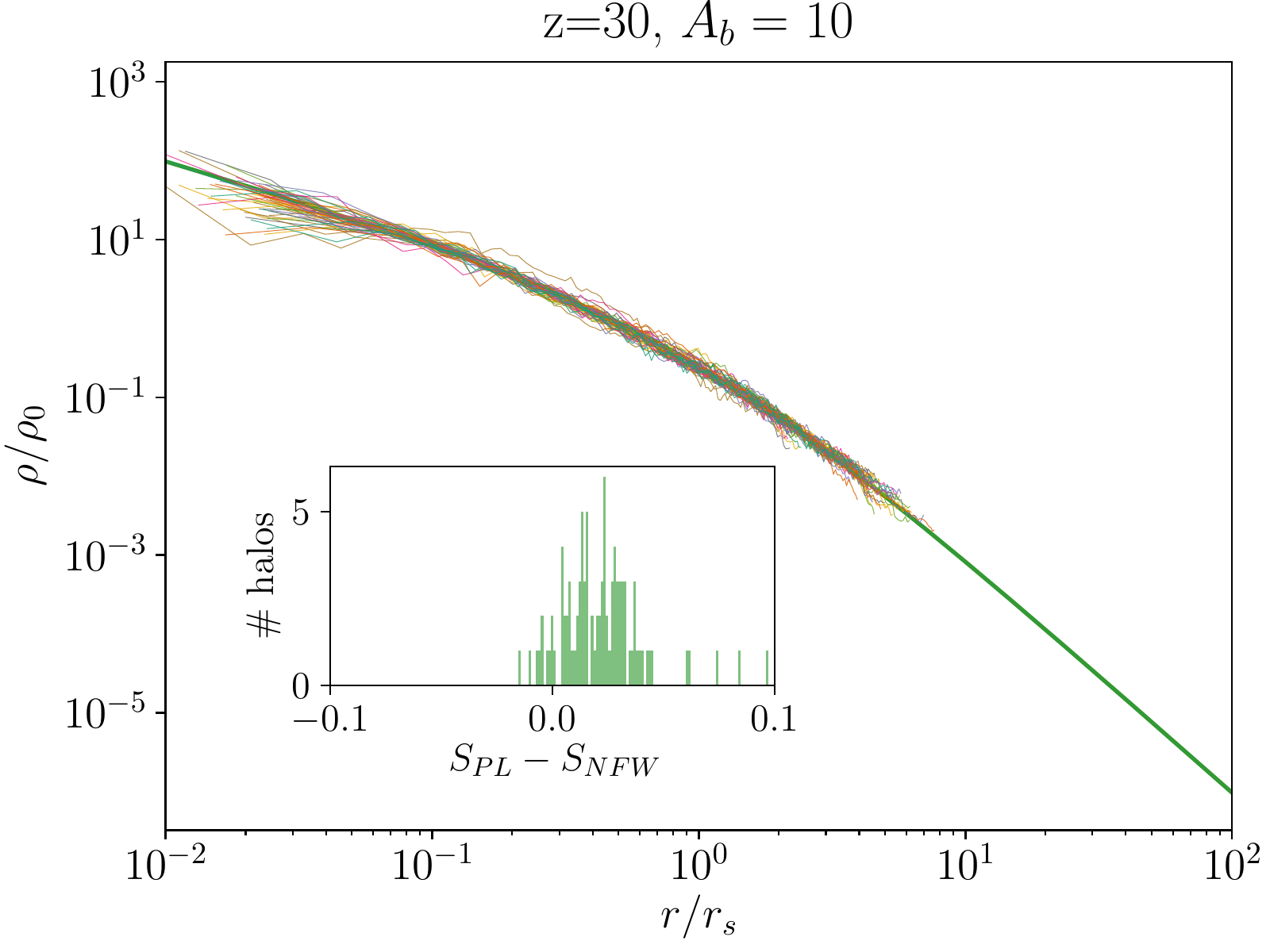}&
    \includegraphics[width=0.31\linewidth]{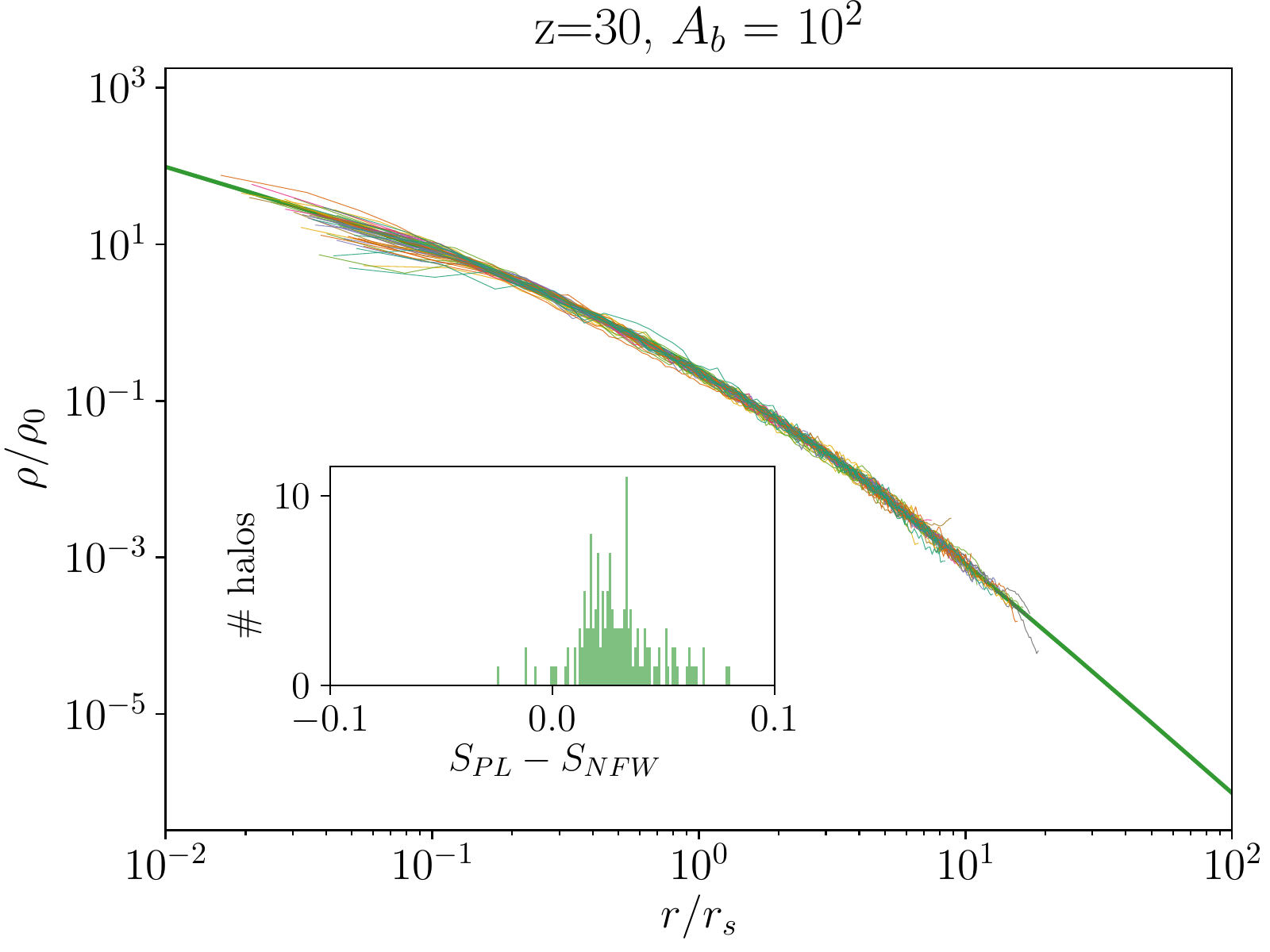}
\end{tabular}
\caption{\label{fig:seed1794profilesNFW1} (Color online) The evolution of halos' profiles with the increasing amount of boost. The NFW analytical prediction is shown in green. Histograms show the difference of a measure for goodness of fit (equation \eqref{eq:fittingstat}) between the power-law and NFW. Almost all halos seem to be a better fit to NFW than to power-law.}
\end{figure*}

\begin{figure*}[tb]
\centering
\begin{tabular}{ccc}
    \includegraphics[width=0.31\linewidth]{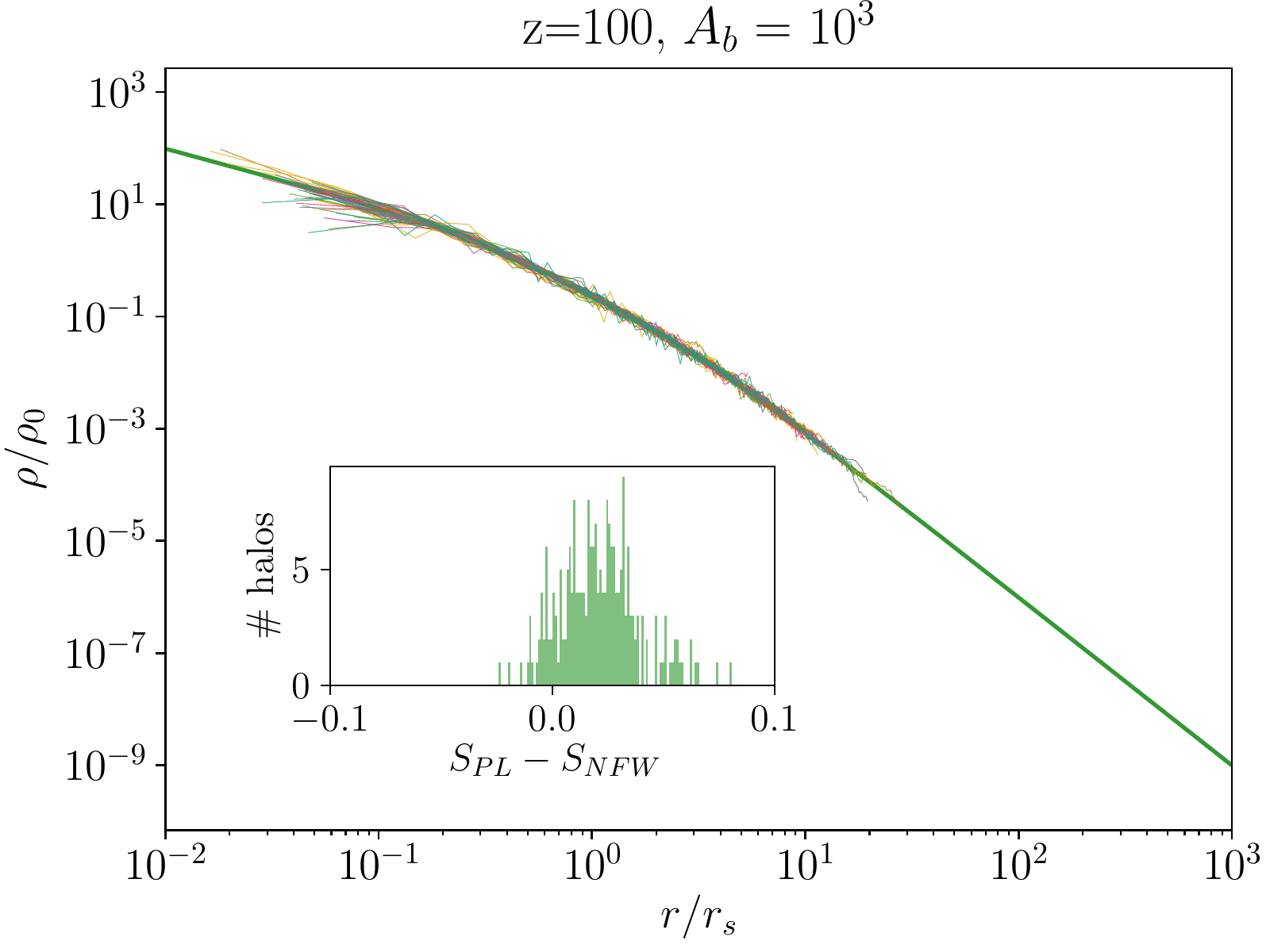}&
    \includegraphics[width=0.31\linewidth]{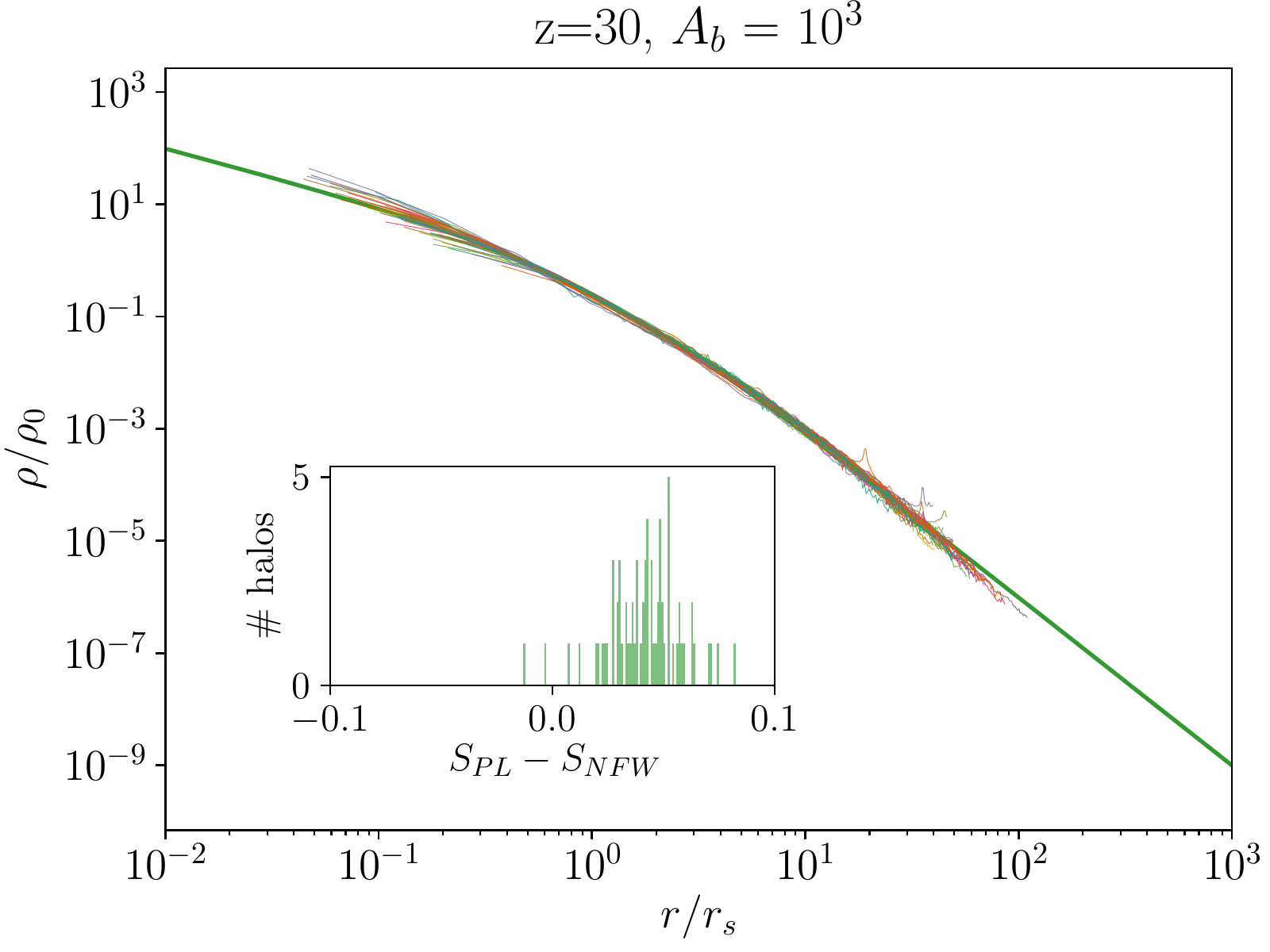}&
    \includegraphics[width=0.31\linewidth]{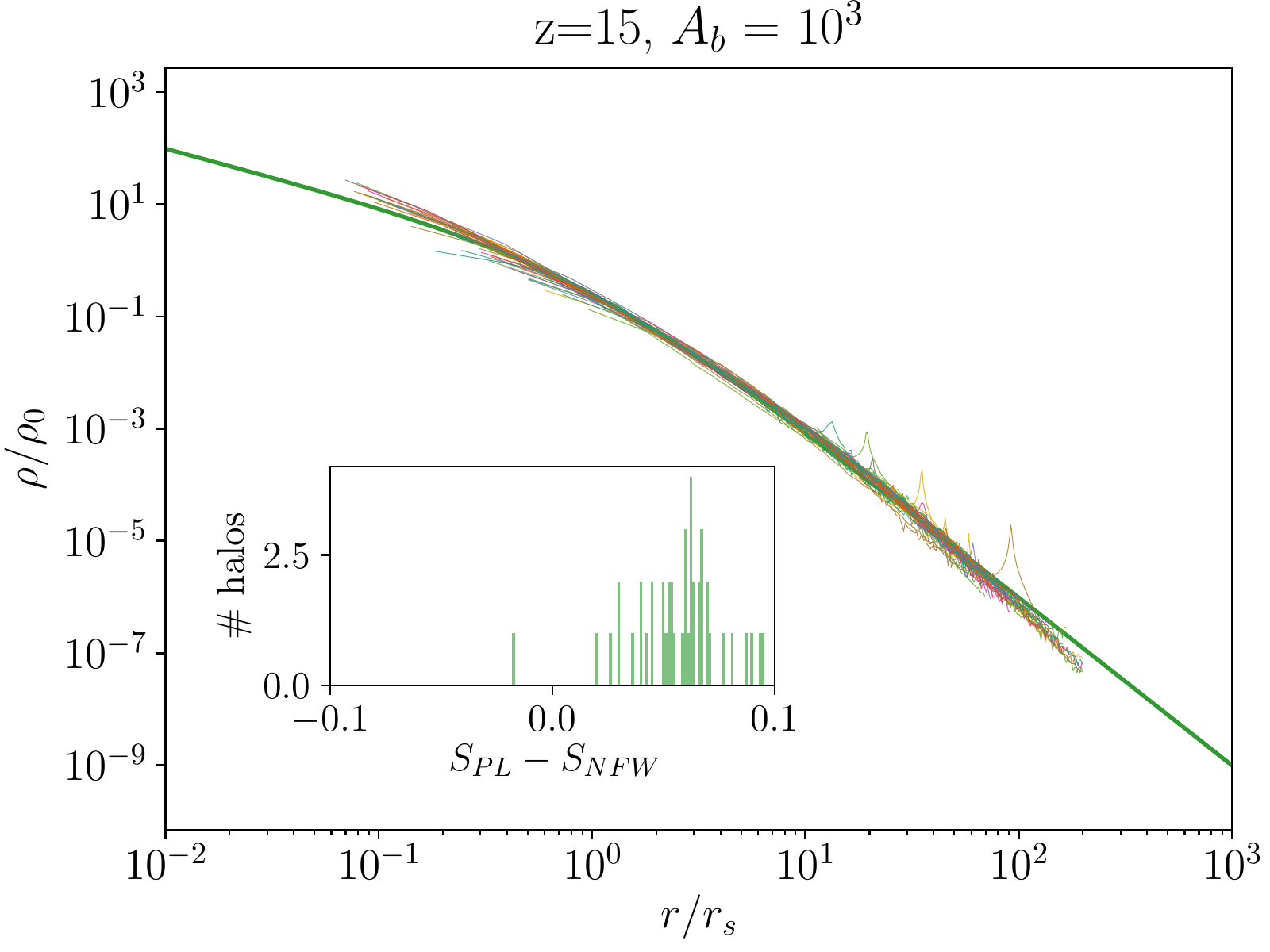}
\end{tabular}
\caption{\label{fig:seed1794profilesNFW2} (Color online) The rescaled profiles for $A_b=10^3$ at $z=100$, $30$ and $15$. From the positions of the lines on top of the NFW reference line it is clear that $r_s$ is getting smaller as time increases. As with Fig. \ref{fig:seed1794profilesNFW1} the NFW profile is a better fit than a power-law for almost all of the profiles.}
\end{figure*}

As might be expected, given our results from section \ref{sec:spherical-overdensity}, we find that most of the halos fit NFW profiles better than power-law (see Figs. \ref{fig:seed1794profilesNFW1} and \ref{fig:seed1794profilesNFW2}). Moreover, there are no obvious exceptional halos, even when we set $A_b=10^3$. There are a few halos that fit the power-law better than NFW; however none stand out as much as the special halo in the peak-to-background 15 simulation in section \ref{sec:snr15} (see, for example, the red line in Fig. \ref{fig:snr15profilesNFW} for $z=30$, $15$ and $10$). We expect this better fit arises here simply because our resolution is not good enough to resolve the $r_s$ of the NFW profile for these \emph{lower mass} halos. A similar effect was seen for the non-special halos in the previous section.

\begin{figure*}[tb]
\centering
\begin{tabular}{ccc}
    \includegraphics[width=0.31\linewidth]{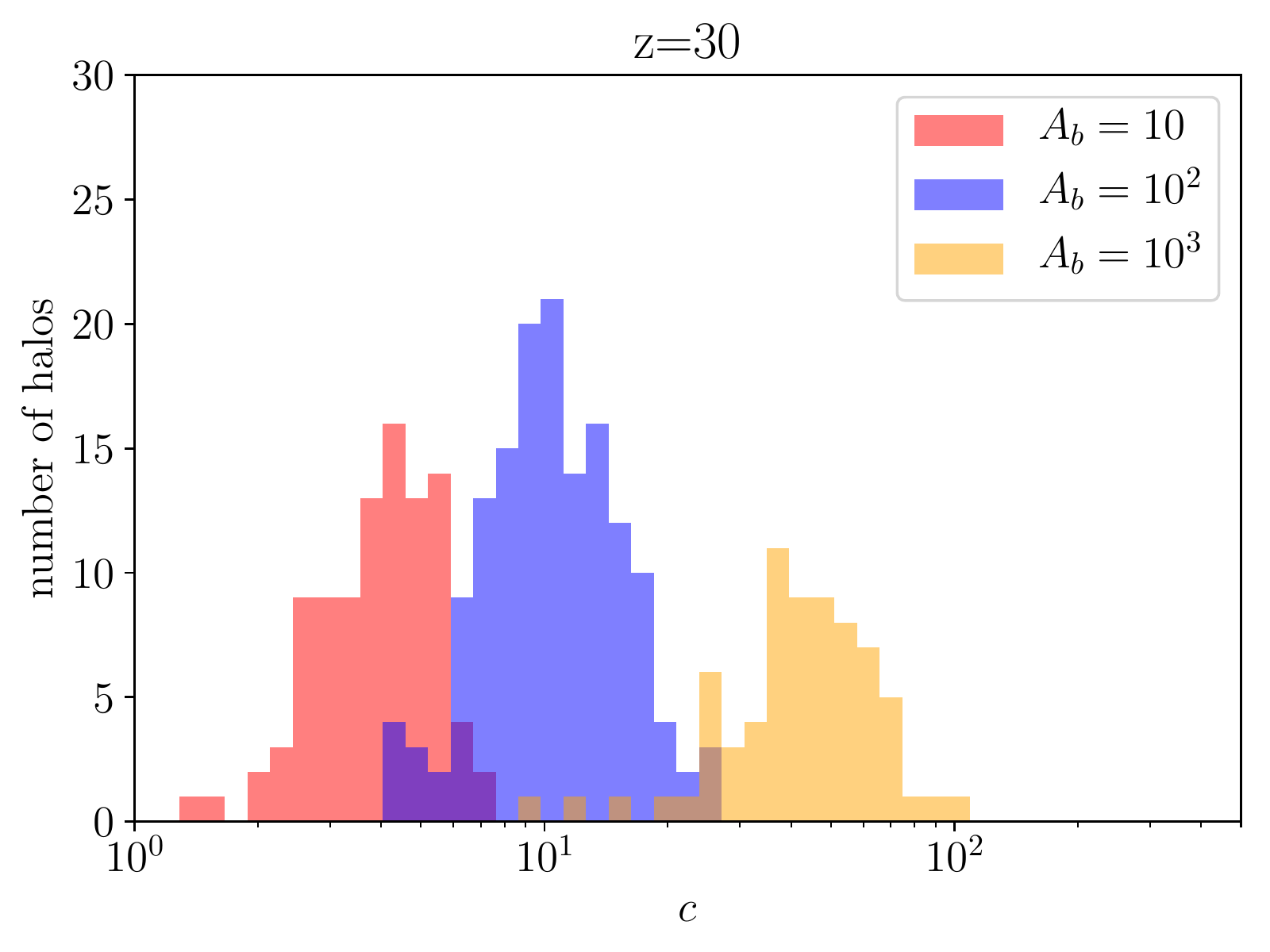}&
    \includegraphics[width=0.31\linewidth]{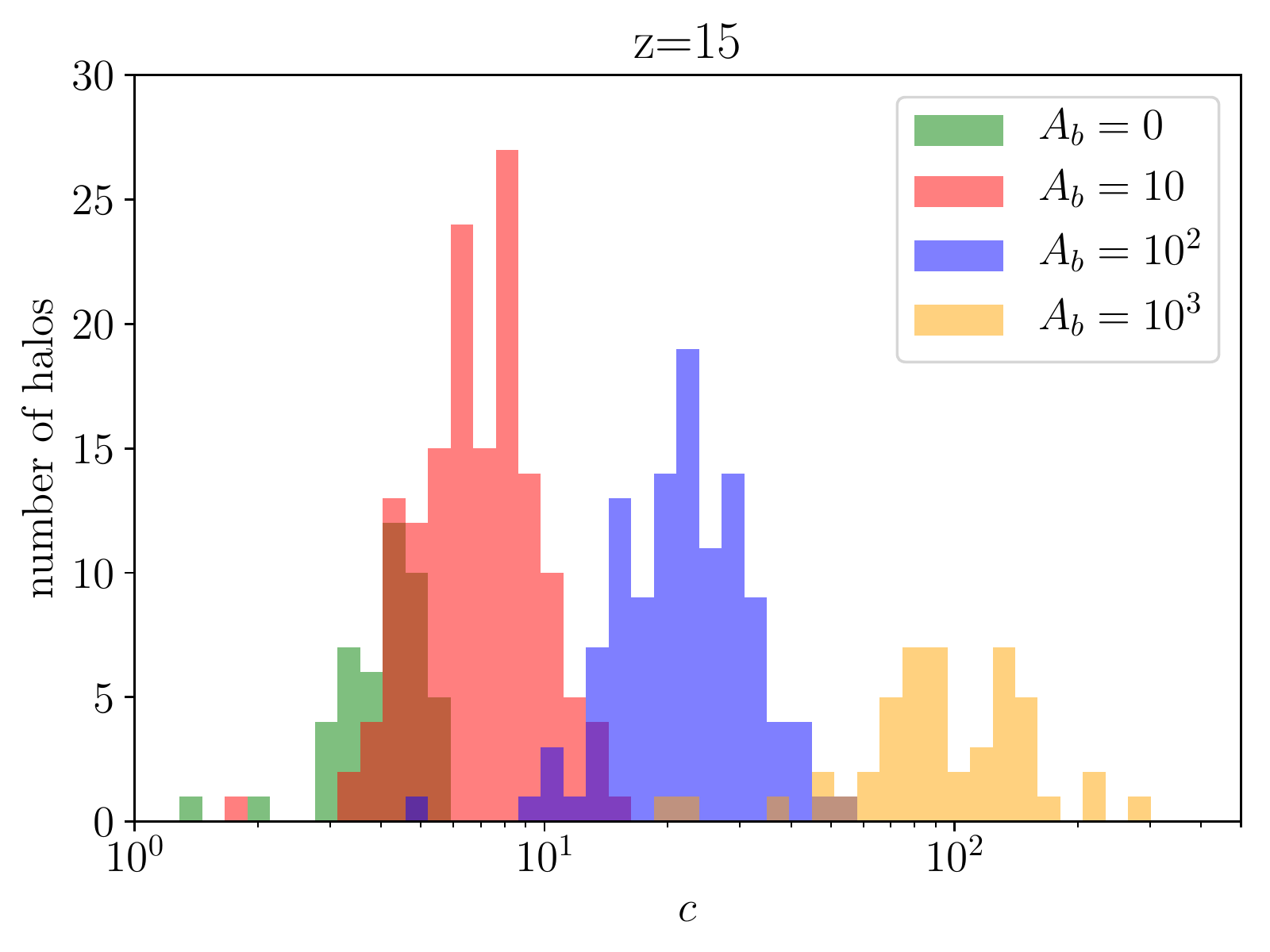}&
    \includegraphics[width=0.31\linewidth]{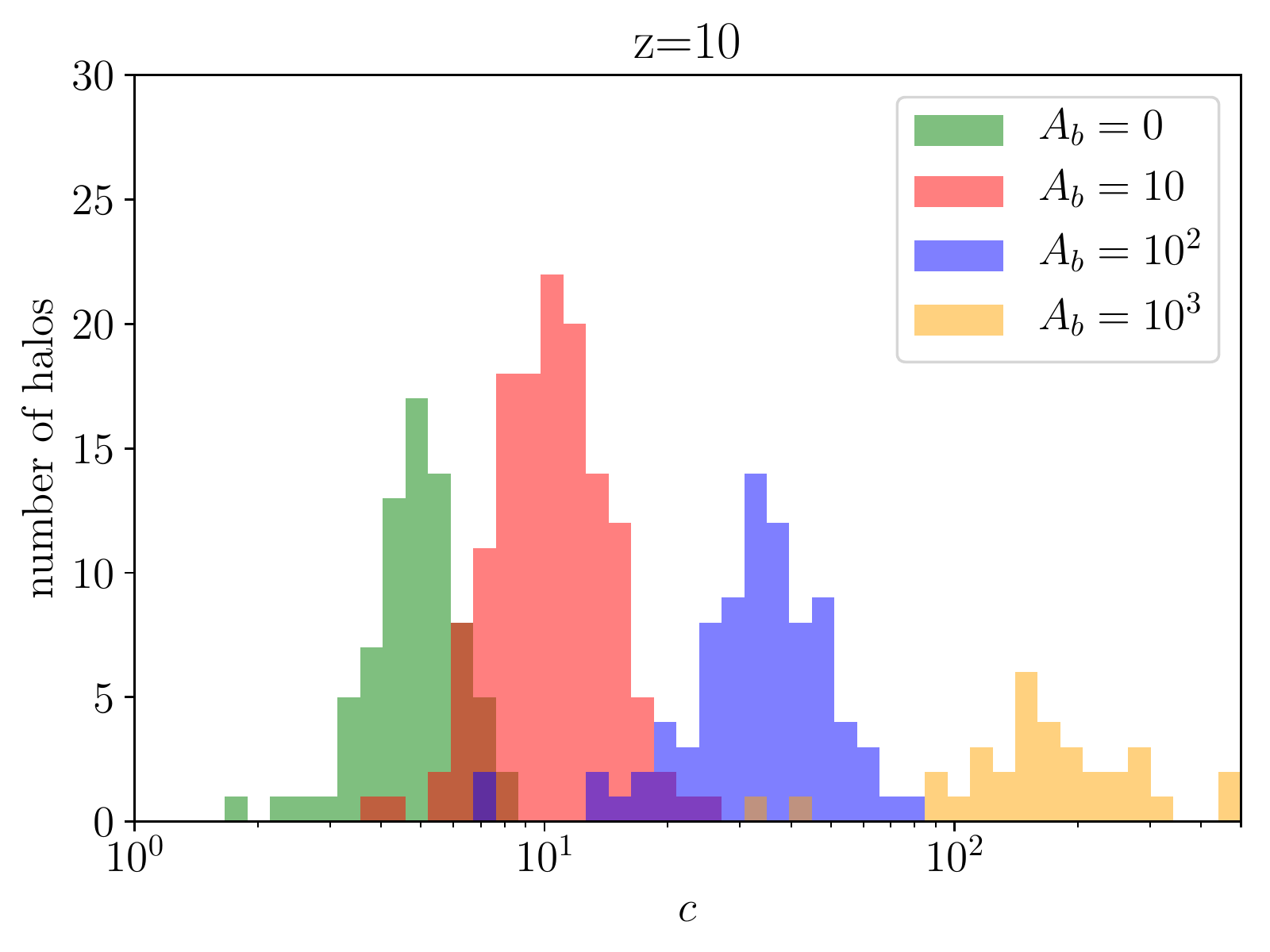}\\[3\tabcolsep]
\end{tabular}
\caption{\label{fig:c-histogram} (Color online) Concentration parameter $c$ for the halos in simulations where we boosted the power spectrum. The halos in higher-boost simulations are much more compact. Concentration also grows with time. We only retain halos with $M_\vir>10^2 M_\odot/h$. Note that at $z=15$ there already exist halos in the $A_b=10^3$ simulation with $c>200$.}
\end{figure*}

Histograms of the concentration parameter $c$ are shown in Fig.~\ref{fig:c-histogram}. At redshift $z=30$, concentration parameters of $c\gtrsim100$ can occur for the $A_b=10^3$ simulation. Given that we expect this to grow with time, the concentration today would be much larger. This shows that although there may be no UCMH candidates in our simulations the structures that do form are still much more compact than those in a simulation without a boosted power spectrum. It is worth stressing again that when the input power spectrum is boosted \emph{all} the structures that form are much more compact. It is not just the rare, extreme fluctuations that experience this effect. This is because the boosted initial conditions form structures earlier, when the whole universe was more dense. Therefore these structures also virialise at much larger densities and remain much denser at later times.

A similar, but less pronounced effect is also seen for the smaller boosts.

\section{UCMH observability and WIMP annihilation}\label{sec:WIMP}

Numerous observational tests have been either proposed or enacted to constrain the number density of UCMHs, and these observational 
limits translate into upper bounds on the primordial power spectrum. These constraints assumed a dense $r^{-9/4}$ inner profile of the 
halos, which we have shown to be unlikely to form. In this section we briefly calculate how the observability of these halos changes when using 
an NFW halo with a large concentration parameter, of the form observed in our simulations. Whilst calculating the constraint on the power 
spectrum goes beyond the scope of this work, we use our simulation results in this section to show how the expected signature of WIMP 
annihilation in the dense center of the halos reduces when using a realistic halo density profile. 

The expected gamma-ray flux from WIMP annihilation within a halo at distance $d$ from the Earth is given by \cite{Josan:2010vn}
\begin{equation}
\Phi_\gamma=\frac{\Phi_{\rm astro}\Phi_{\rm particle}}{2 d^2}
\end{equation}
where $\Phi_{\rm particle}$ depends on the WIMP particle mass and annihilation cross section, which is assumed to be independent of the halo properties, e.g.~the velocity distribution of the particles (see \cite{Bringmann:2012ez} for a justification). 

The astrophysical part is given by an integral of the density squared
\begin{equation}
\Phi_{\rm astro}=\int_0^{r_h} \rho(r)^2 r^2 {\rm d} r
\label{eq:Phi-astro}
\end{equation}
where $r_h$ is the radius of the halo, which we always take to be $r_{\rm vir}$. In practice, although the majority of the NFW halo mass is within the outer part of the profile where $\rho\propto 1/r^3$, the annihilation signal is dominated by the innermost part with the greatest density, which satisfies $r<r_s$ so it does not matter much where we cut-off the integral. Note that the physical radius and density must be used.

For halos with an $r^{-9/4}$ profile, the signal is dominated by the central density, which must be cut off at some maximum value. Bringmann et al.\ estimate a maximum possible density of the UCMHs today of
\begin{equation}\label{rhomax}
\rho_{\rm max}\simeq \frac{m_\chi}{\langle\sigma v\rangle (t_0-t_i)} = K \rho_{c,m},
\end{equation}
where 
\begin{equation}
K\simeq5\times10^{16},
\end{equation}
calculated assuming fiducial values of the WIMP mass $m_\chi=1\,{\rm TeV}$, thermally averaged cross section $\langle\sigma v\rangle=3\times 10^{-26} {\rm cm}^3 {\rm s}^{-1}$, and the age of the Universe $t_0=13.7\,{\rm Gyr}$. The UCMH formation time $t_i$ is irrelevant provided that $t_i\ll t_0$ \cite{Bringmann:2012ez}.
The critical density of the Universe today is $\rho_c=415 M_\odot h^2/{\rm kpc}^3$, and 
$\rho_{c,m}\equiv\Omega_m \rho_c$. The UCMH profile 
is given by
\begin{equation}
\rho_{\rm UCMH} =\begin{cases}
    \rho_{\rm max} & \text{if $r<\rc$},\\
    \rho_{\rm max}\left(\frac{r}{\rc}\right)^{-\frac94} & \text{if $r>\rc$}.
  \end{cases}
\end{equation}

\begin{figure}[t]
\includegraphics[width=\columnwidth]{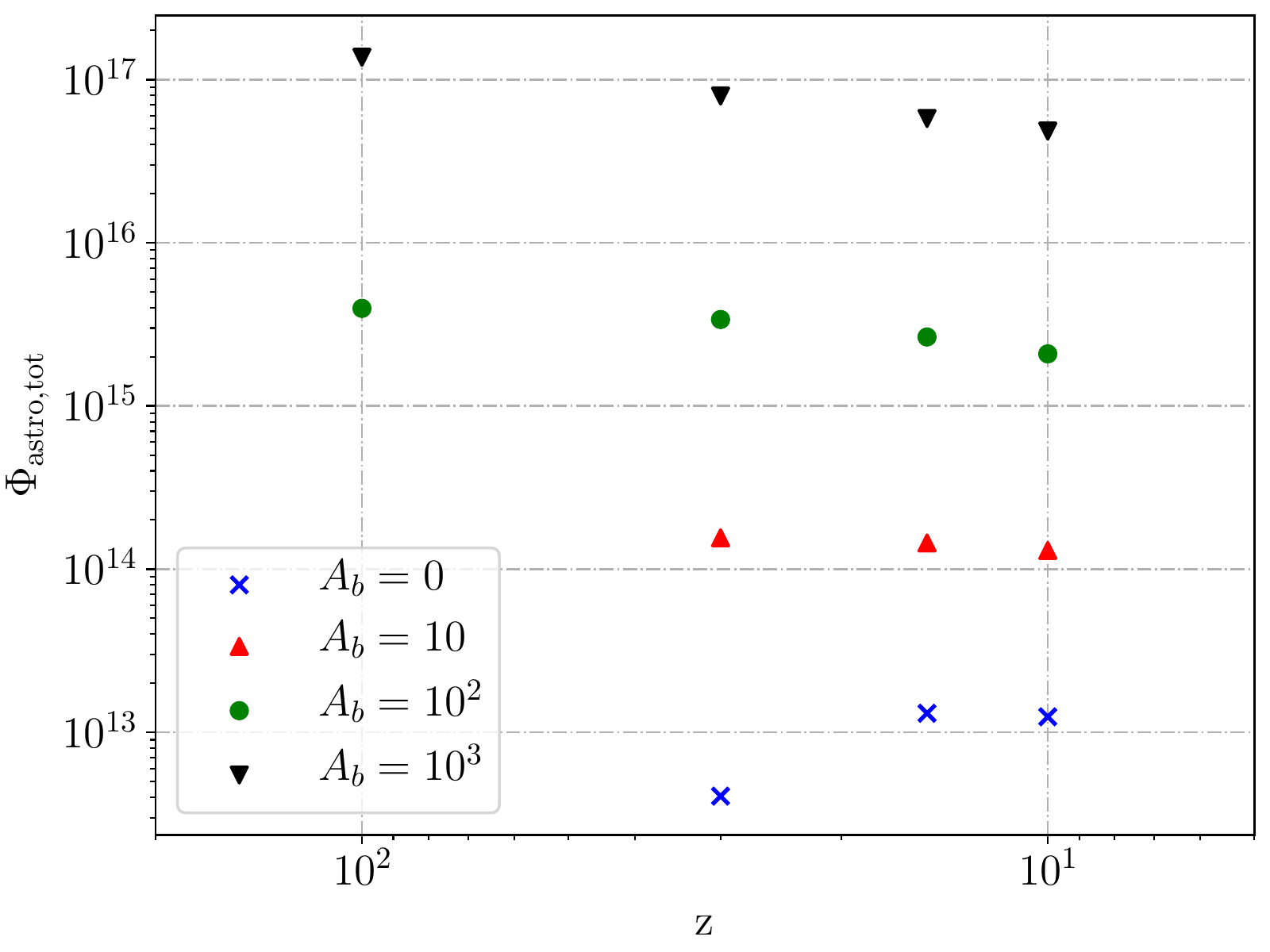}
\caption{\label{fig:WIMPsignal} (Color online) The total $\Phi_{\rm astro}$ part of the WIMP-annihilation signal from the most massive halos plotted against $z$ for the power spectrum boosted by 4 different amounts. Notice how the strength of the signal typically decreases slowly with redshift, which we caution may be a numerical artifact, see the text after Eq.~\eqref{eq:Phi-NFW}. To calculate this signal, we take into account all the halos with $M_{\rm vir} \gtrsim 3\,M_{\odot}$. For the two smallest boosts at $z=100$, no such halos are identified in the simulation.}
\end{figure}

Using the definition of the virial mass in terms of the virial radius (which is independent of the density profile)
\begin{equation}
M_{\rm virial}=\frac{4\pi}{3}178\rho_{c,m} r_{\rm vir}^3,
\end{equation}
where $\rho_{c,m}$ is the critical density of matter (of both CDM and baryons, because our simulations treat them equally). 
Calculating the density contrast centered on a UCMH to a radius $r_{\rm vir}$
\begin{equation}
\begin{split}
1+\delta_{\rm UCMH}(r_h) &=\frac{1}{\rho_{c,m}} \frac{3}{r_{\rm vir}^3} \int_0^{r_{\rm vir}}\rho_{\rm UCMH}(r) r^2 dr  \\
&\simeq 4K\left(\frac{\rc}{r_{\rm vir}}\right)^{9/4},
\end{split}
\end{equation}
leads to 
\begin{equation}
\frac{\rc}{r_{\rm vir}} \simeq  \left(\frac{179}{4 K}\right)^{4/9}\ll1.
\end{equation}
Using the above results, we find the WIMP-annihilation signal is 
\begin{equation}
\Phi_{\rm astro,UCMH} \simeq \rc^3 \rho_{\max}^2=K^{2/3}\left(\frac{179}{4}\right)^{4/3} r_{\rm vir}^3 \rho_{c,m}^2.
\end{equation}
For an NFW profile, we can similarly evaluate (\ref{eq:Phi-astro}) to calculate the WIMP-annihilation signal from such a halo, and then compare it to the UCMH result for a halo with the same mass (and hence the same $r_{\rm vir}$),\footnote{Although the density of the NFW profile also diverges at the center, the impact of including the maximum density restriction given by \eqref{rhomax} is negligible.}
\begin{equation}
\begin{split}
\Phi_{\rm astro,NFW}&= r_s^3\rho_0^2 \frac{c(3+c(3+c))}{3(1+c)^3}  \\
 &= 1.2\times 10^3  r_{\rm vir}^3 \rho_{c,m}^2 \frac{c^4(3+c(3+c))}{(1+c)^3}  \\
 &\times \frac{1}{\left(\log(1+c)-c/(1+c)\right)^2},
\end{split} \label{eq:Phi-NFW}
\end{equation}
where we note that the result strongly depends on the concentration parameter, $c=r_{\rm vir}/r_s$, with an approximate $c^3$ dependence in 
the limit of $c\gg1$. For this reason the large values of $c$ generated by boosting the power spectrum do give rise to much larger WIMP 
annihilation signals than would be the case with an unboosted power spectrum for halos of the same mass and also with an 
NFW profile. We show the mild redshift evolution of the WIMP-annihilation signal in Fig.~\ref{fig:WIMPsignal} for all four levels of boost. Increasing $A_b$ by an order of magnitude has a much larger effect on $\Phi_{\rm astro,tot}$, the total value of $\Phi_{\rm astro}$ added up for all halos, than the redshift evolution of any given boost with redshift. Although the WIMP-annihilation signal does initially increase with redshift for the unboosted simulation, we expect this is a numerical artefact of many protohalos having not yet reached sufficient size to be detected by \textit{ROCKSTAR}. Similarly, we caution that the slight decrease in $\Phi_{\rm astro,tot}$ with redshift may be due to the decreasing minimum physical length we can resolve in an expanding background, meaning that we resolves the cores of the halos less well at late times.

We calculated the WIMP-annihlation signal by integrating (\ref{eq:Phi-astro}) which assumes spherical symmetry, and compared the result to (\ref{eq:Phi-NFW}) using the best fit values 
of $c$ from Sec.~\ref{sec:NFW-fit} and $r_{\rm vir}$ outputs from \textit{ROCKSTAR}, finding the results agree within a factor of 2.  
We note that Kohri et al.\ \cite{Kohri:2014lza} have previously shown that the WIMP-annihilation signal is extremely sensitive to the exponent 
$\alpha$ when they assumed a truncated power law profile of the form $\rho\propto (1+r/r_{\rm cut})^\alpha$.

However, even for the largest values of $c$ observed (using a boost factor $A_b = 1000$), the WIMP-annihilation signal as far as we are able to resolve it remains about a factor of $10^4$ smaller than it 
would be for a UCMH with the same mass, which means they would have to be 100 times closer in order to be equally observable. Hence we 
can only observationally rule out their existence in a volume one millionth as large as could be probed for halos with $r^{-9/4}$ profiles.
The NFW halos are however much more common than the assumed abundance of UCMHs given the same initial conditions, meaning that the observational constraints on the power 
spectrum may not weaken as strongly as may be expected, but a detailed study of this issue goes beyond the scope of this paper. 

\begin{figure*}[tb]
\centering
\begin{tabular}{cc}
    \includegraphics[width=0.47\linewidth]{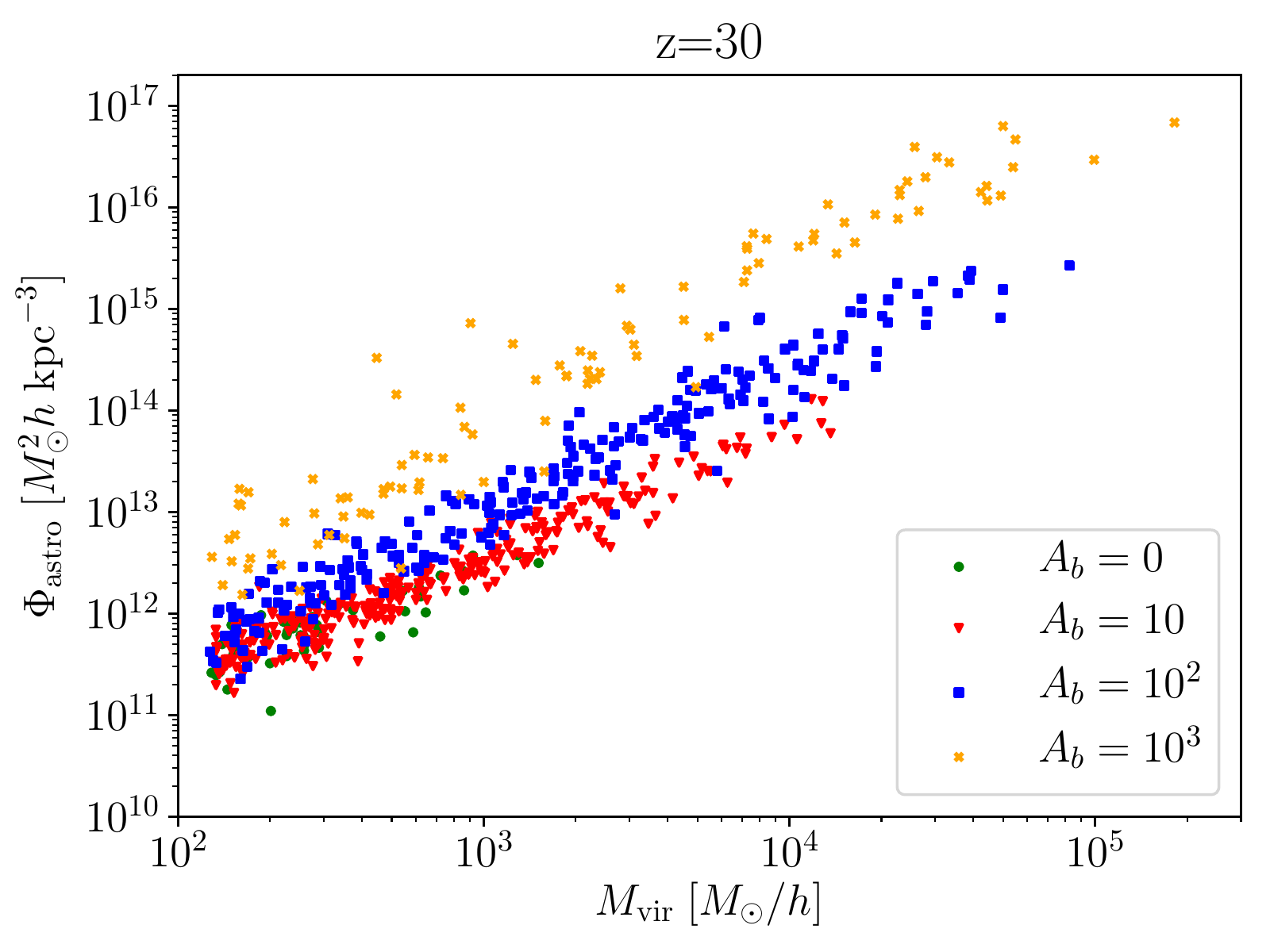}&
    \includegraphics[width=0.47\linewidth]{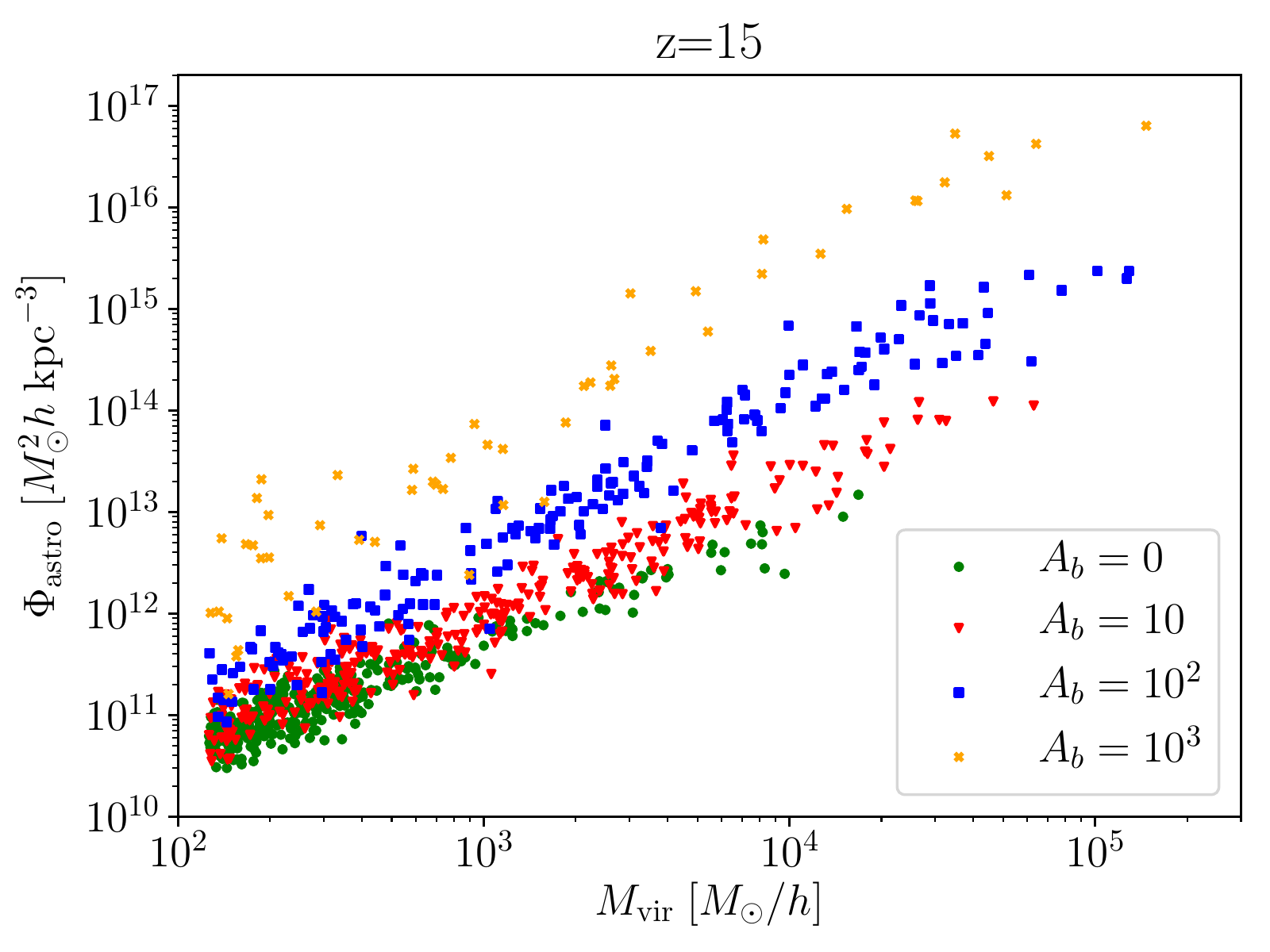}\\[2\tabcolsep]
   \includegraphics[width=0.47\linewidth]{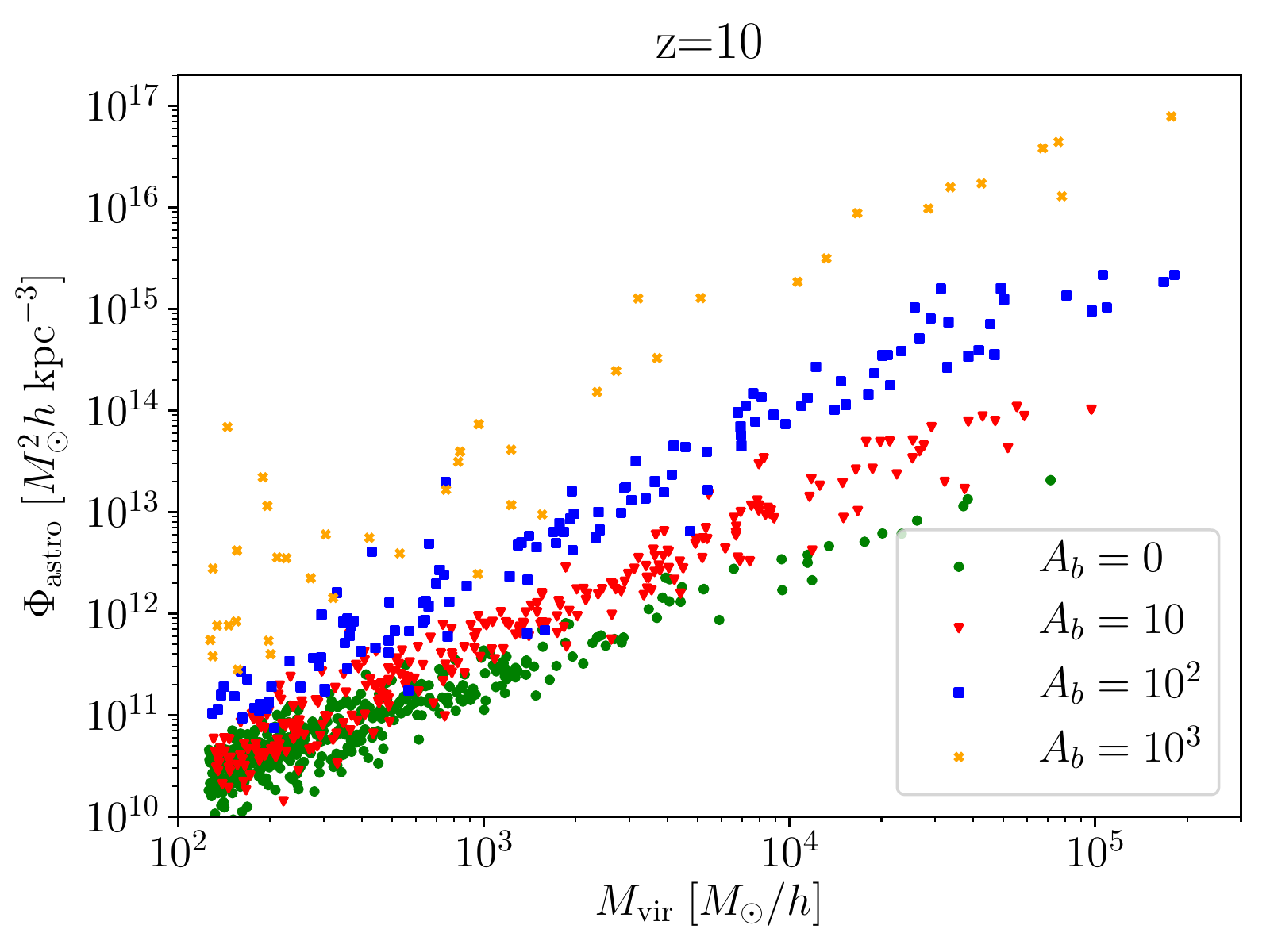}&
    \includegraphics[width=0.47\linewidth]{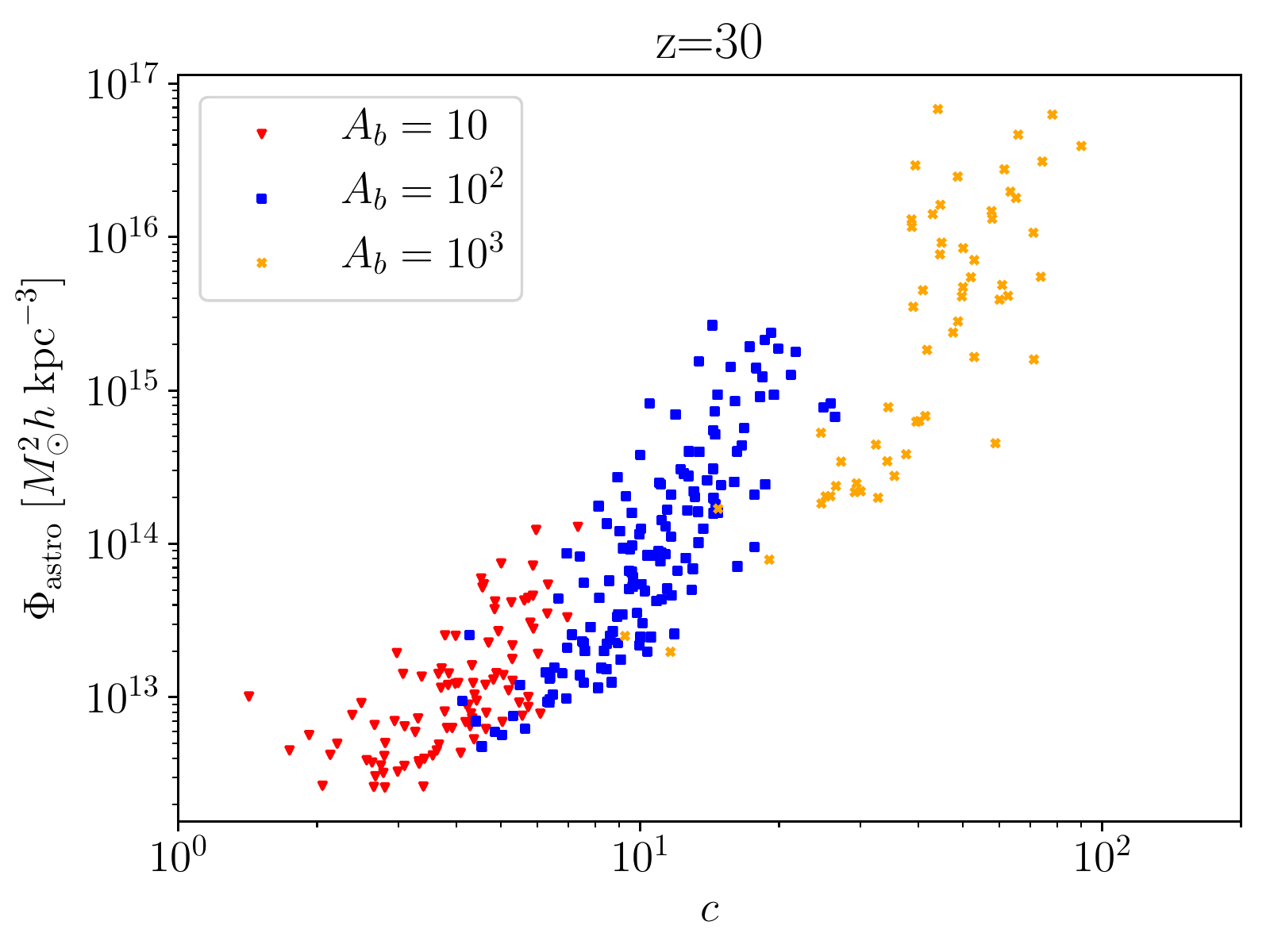}\\[2\tabcolsep]
\end{tabular}
\caption{\label{fig:Mvir-c-WIMP-scatter} (Color online) Scatter plots for the WIMP-annihilation signal plotted against $M_{\vir}$ for three different redshifts (top and left panels) and against $c$ for $z=30$ (bottom right). Different amplitudes of the boost are represented with different color. It is curious that the WIMP-annihilation signal appears to depend on concentration in a way that doesn't depend on the size of the boost. However, there is a lot of scatter in this relationship.
}
\end{figure*}

In Fig.~\ref{fig:Mvir-c-WIMP-scatter} we explore whether there is any connection between the combinations of three different parameters: the virial mass of a
halo $M_{\vir}$, its concentration parameter $c$, and a measure for the WIMP annihilation signal. The WIMP-signal measure is higher for heavier halos and more compact ones. It is also significantly higher in simulations with a higher boost. 

\begin{figure*}[bt]
\begin{tabular}{cc}
\includegraphics[width=\columnwidth]{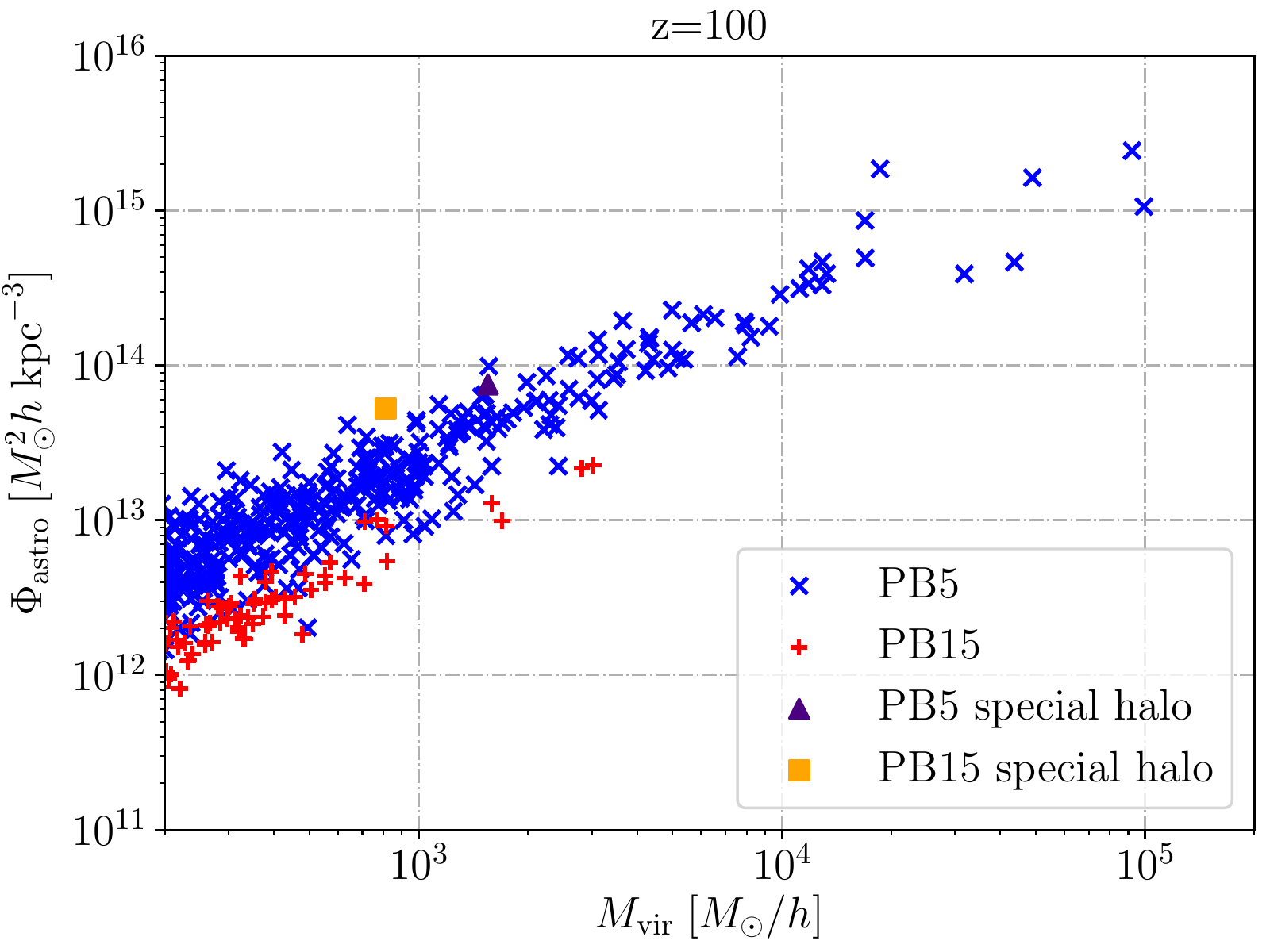}
\includegraphics[width=\columnwidth]{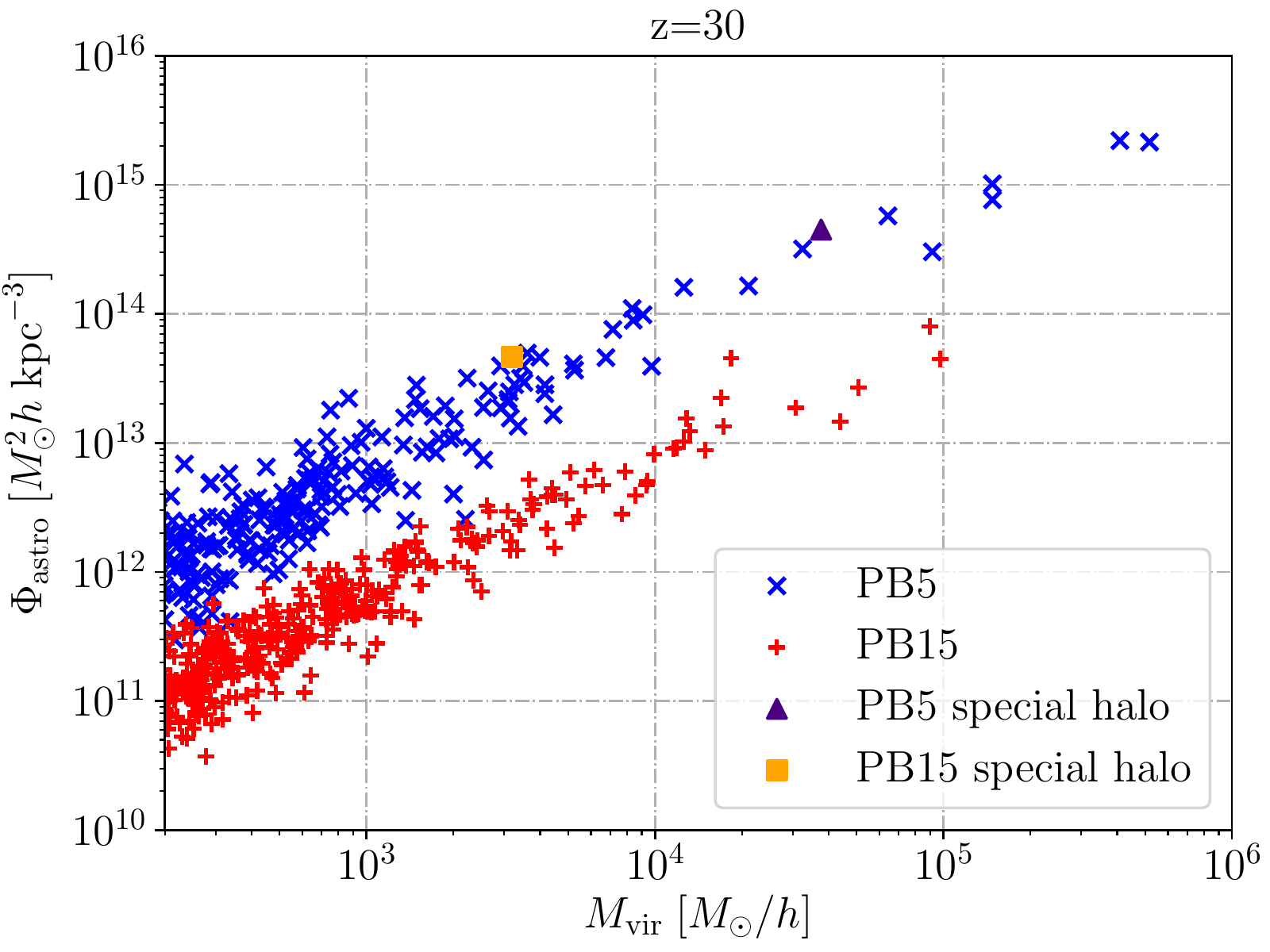}
\end{tabular}
\caption{\label{fig:WIMP-pb-z30} (Color online) The $\Phi_{\rm astro}$ part of the WIMP-annihilation signal from all halos with $M_{\rm vir}> 2\times10^2 M_\odot /h$. For the peak-to-background ratio 15 simulations, the value of $\Phi_{\rm astro}$ is significantly boosted compared to all other halos with the same mass from the same simulation, due to the steeper profile density, as shown by 
Fig.~\ref{fig:pbr15alphavsmass} for the same two redshifts. However for the peak-to-background 5 simulation, we see that the special seed no longer has a larger value than average, but that its mass is considerably larger than in the peak-to-background 15 simulation, especially at $z=30$, meaning that the special seed has 
has undergone more merging.
}
\end{figure*}

In Fig.~\ref{fig:WIMP-pb-z30} we show the WIMP-annihilation signal as a function of virial mass of all halos. The most interesting feature is that the WIMP-annihilation signal of the special seed in the peak-to-background 15 simulation is an order-of-magnitude larger than any of the other halos in the same simulation with comparable mass. If we had better resolution, we expect that the WIMP-annihilation signal from this halo would become even larger than what is plotted. The integral of the density squared is dominated by the very center of the halo where the density is largest, and this halo has a steeper profile towards the center than all of the others, see Fig.~\ref{fig:snr15profiles}.

\section{Conclusions}
\label{sec:conclusions}

We have performed the first 3D N-body simulations of UCMH formation. Starting with an isolated spherical overdensity we have shown that this
would form a steep power-law profile with the density scaling close to $\rho\propto r^{-9/4}$ \cite{1985ApJS...58...39B}, and we are
able to resolve this profile by up to 3 orders of magnitude of length scales at low redshift. When we include 
random density fluctuations with an amplitude typically 15 times smaller than the special halo, 
we observe that the halo descended from the special seed becomes somewhat disrupted, with the density profile becoming shallower towards the center (compare Figs.~\ref{fig:gaussian-snrinf} and \ref{fig:snr15profiles}). 
This flattening occurs on such small scales that it does not make a significant difference to the numerical values of the halo properties we extract, such as the power-law steepness $\alpha$ or the WIMP-annihilation signal. Since the WIMP-annihilation signal is dominated by the square of the density in the core of the halo, our estimates for the WIMP-annihilation signal
present only the lower bound on this quantity, due to the limitations of our resolution.

However, even though the special halo flattens a little it still remains exceptional relative to the other halos that form in the simulation box. Its steepness and WIMP-annihilation signal are both much larger than any other comparable mass halo in the simulation (see Figs. \ref{fig:pbr15alphavsmass} and \ref{fig:WIMPsignal}). It is therefore likely that the inner part of this halo, unresolvable by our simulations, remains close to a power-law.

When we further increase the size of background fluctuations to be five times smaller than the special halo we see the special halo become further disrupted. In fact, in this simulation, even for the special halo, we find an NFW profile is a better fit than a power-law (see Fig. \ref{fig:snr5profilesNFW}). Moreover, although it starts as equivalent to a $5$-$\sigma$ fluctuation of the background it has ceased to be exceptional even by $z=300$. These two facts strongly suggest that even below the scales we can resolve the special halo will not have a steep UCMH-like profile.

Note that in all of these situations the initial special seed remains the same size, it is only the background fluctuations that change. Therefore, we can conclude that in order to form this UCMH-like power-law profile an initial fluctuation must be substantially larger than five times its background.  
Ricotti and Gould  estimated that the critical density threshold for collapse into a UCMH was $\delta_c=10^{-3}$ \cite{Ricotti:2009bs} and this estimate was refined to include a scale dependence by Bringmann et al.~\cite{Bringmann:2011ut}. We have here shown that in realistic cases $\delta_c$ is strongly dependent on the environment.
The subsequent growth, profile steepness and WIMP-annihilation signal are all affected by the neighbouring perturbations.

We note that although almost all constraints from UCMHs have been made assuming a steep $r^{-9/4}$ profile to hold down to extremely small radii, the expectation from N-body simulations is that the emergence of NFW profiles is generic due to radial instabilities which will always exist and grow within realistic simulations \cite{Angulo:2016qof,Ogiya:2017hbr,2010arXiv1010.2539D}. However, we perform the first 3D simulations with boosted initial power spectra, probing small scales and starting deep in the radiation dominated era. We confirm that the NFW profile remains generic. In section \ref{sec:extreme} we have estimated an upper bound on the typical distance between an extremely large overdensity and the nearest overdensity with a comparable size, and shown that even a rare $5$-$\sigma$ fluctuation will not be isolated from the evolution of neighbouring halos over a long period of time.

Despite providing evidence that UCMHs of the form typically considered in the literature are not very likely to be realised in nature, we stress that our simulations with 
the initial power spectrum boosted on the $\kpc$ scale
do form a significant number 
of dense NFW-like halos with masses around $10^4-10^5 M_\odot/h$. 
Furthermore, these halos have
extremely large values of the concentration parameter $c=r_{\rm vir}/r_s$, which can grow to over 100 by redshift ten.  Assuming the central density remains constant and the concentration grows roughly as $c\propto 1/(1+z)$ \citep{Bullock:1999he}
this would imply concentrations as large as 1000 by redshift zero. Subsequently, it would be very interesting to extend our simulations to redshift zero to study the stability of the halos we simulate with large concentration parameters. However, this would also require the inclusion of baryonic effects on these extremely small scales.

Given that the 
presence of halos with profiles close to $r^{-9/4}$ is unrealistic in the 
Universe the observational upper bounds on the primordial power spectrum on small scales, 
derived from the non-observation of UCMHs will have to be re-evaluated. {However, given that we have also shown that when the primordial power spectrum is boosted even \emph{typical} halos form at much higher densities and are thus much more compact than usual there are likely to be other means to constrain these small scales, which remains to be explored.

\acknowledgments

We gratefully acknowledge Camilla Penzo for providing her version of \textit{RAMSES} that includes the contribution of radiation density in the Hubble function. The authors thank Jens Chluba, Peter Coles, Neal Dalal, Oliver Hahn, Massimo Ricotti, Patrick Scott, David Seery and Teruaki Suyama for interesting discussions.
JA, CB and MG would like to thank the Department of Physics at the University of Auckland for kind hospitality. 
SH thanks Sussex University for hosting him on a return visit. CB is supported by a Royal Society University Research Fellowship.
This work was supported by a grant from the Swiss National Supercomputing Centre (CSCS) under project ID s710.

\appendix

\section{Notes about the simulations and halo analysis}
\label{sec:app}

\subsection{Initial conditions and quasilinear evolution}

In order to set up our initial conditions we use a customized version of the initial condition generator
that is implemented in \textit{gevolution} \cite{Adamek:2015eda}. In the public version of \textit{gevolution}, the initial conditions are set up in terms
of a linear displacement field and velocity potential that are obtained by multiplying a Gaussian realization of the primordial curvature perturbation
by the respective linear transfer functions. The latter can be computed using a Boltzmann code such as \textit{CAMB} \cite{Lewis:1999bs} or \textit{CLASS}
\cite{Blas:2011rf}. As usual, the particle initial conditions are obtained by displacing particles from a regular lattice according to the gradient of
the displacement field, and assigning velocities according to the gradient of the velocity potential. In our modification of this algorithm we introduce
the possibility to create an initial configuration corresponding to a spherical Gaussian overdensity that can act as a seed for a UCMH. Using the superposition
principle that applies in the linear regime we allow in general for any linear combination of the random realization and this spherical overdensity. 

The initial conditions are set deep inside the radiation dominated era, at redshift $z = 10\,000$ for the simulations with the special central halo or $z = 5\,000\,000$ for the simulations with the boosted power spectrum. We ignore the
fact that the baryonic component of matter is strongly coupled at that time, assuming that at the small scales relevant for our study the perturbations in the
baryon-photon plasma are irrelevant and we only have to consider CDM. Using \textit{gevolution} we follow the initial logarithmic growth of the matter
perturbations up to the point where the density contrast approaches unity for the most extreme initial conditions. This happens around $z \simeq 3000$.

\subsection{Nonlinear evolution}

The particle-mesh scheme of \textit{gevolution}, which works at fixed spatial resolution, eventually becomes inadequate for tracking the detailed
evolution of the very compact structures we are interested in. One common approach to deal with this issue is to use adaptive mesh refinement (AMR), that is
to successively fine-grain the mesh in regions of high particle density. A cell-based AMR algorithm is implemented in the public code \textit{RAMSES}
\cite{Teyssier:2001cp}. We use a slightly modified version where the Hubble function takes into account radiation density and hence is more accurate at
high redshift. In order to pass from one code to the other we have \textit{gevolution} write a snapshot that we then use as initial data for \textit{RAMSES}.
By running the two codes just a bit further and comparing snapshots at a later time (still before AMR is triggered in \textit{RAMSES}) we convince
ourselves that the change between codes did not introduce any unexpected issues.

For simplicity, we do not use the hydrodynamics modules of \textit{RAMSES} which means that baryons are effectively treated as
dark matter. Baryonic effects are expected to play an important role at low redshift $z \lesssim 10$, but we are not interested in
these aspects here and stop our simulations before they become a serious concern.

\subsection{Spherical shells and smoothing}

Once halos have formed we identify them using the friends-of-friends halo finder algorithm \textit{ROCKSTAR} \cite{Behroozi:2011ju}.
For each halo
the profile of $1 + \delta$ is obtained by counting the number of particles $n$ inside the spherically symmetric shells around the center of the halo and dividing by the volume of each shell:
\begin{equation}
1+\delta=\frac{\rho}{\bar{\rho}} = \frac{n V_{\mathrm{shell}}}{N^3 V},
\end{equation}
where $N^3$ is the number of all particles in the simulation and $V_{\mathrm{shell}}$ and $V$ are the volumes of a chosen shell and the entire box, respectively. To reduce the noise at higher $r$, we have gradually increased the size of the bins such that every bin was larger than the previous by a constant factor: $r_{n+1}=q r_{n}$, which was set to $q=1.01$. Therefore, in the i-th bin the density contrast was 
\begin{equation}
1+\delta=\frac{3n L^3}{4\pi N^3 r_0^3} \frac{(1-q)^3}{(1-q^{i+1})^3-(1-q^i)^3},
\end{equation}
where $L^3$ represents the entire volume of the simulation. For the size of the starting bin $r_0$ we chose four-times the size of a cell for the finest AMR level. In our simulations the maximum refinement level was $16$, and therefore the starting bin was $r_0= 32 /2^{14}\,  \kpc /h$ $=1.95\; \mathrm{pc}/h$.
This approach works as long as the halos are approximately spherical. To avoid empty bins in the shells where there happen to be no particles, we apply Gaussian smoothing with a width small enough to retain the shape of a halo profile. We checked that this does not degrade our resolution.

\subsection{Convergence tests}\label{sec:convergence}
In order to understand up to what minimal radius we can trust our results we ran some convergence tests. In Fig.~\ref{fig:sn15convergence} 
we show two simulations performed with \textit{RAMSES} with the same initial conditions, but different number of particles: $256^3$ and
$512^3$. This corresponds to a mass resolution of $0.169 M_\odot/h$ and $0.021 M_\odot/h$, respectively. In 
both cases the maximum AMR level was $16$. From the plot we conclude that we can trust the lower-resolution simulation down to approximately $r \sim 2 \times 10^{-3} \mathrm{kpc}/h$. Unless indicated otherwise, our numerical results presented in Sections III--V were obtained
with simulations that had $256^3$ particles.

\begin{figure}[h]
\includegraphics[width=\columnwidth]{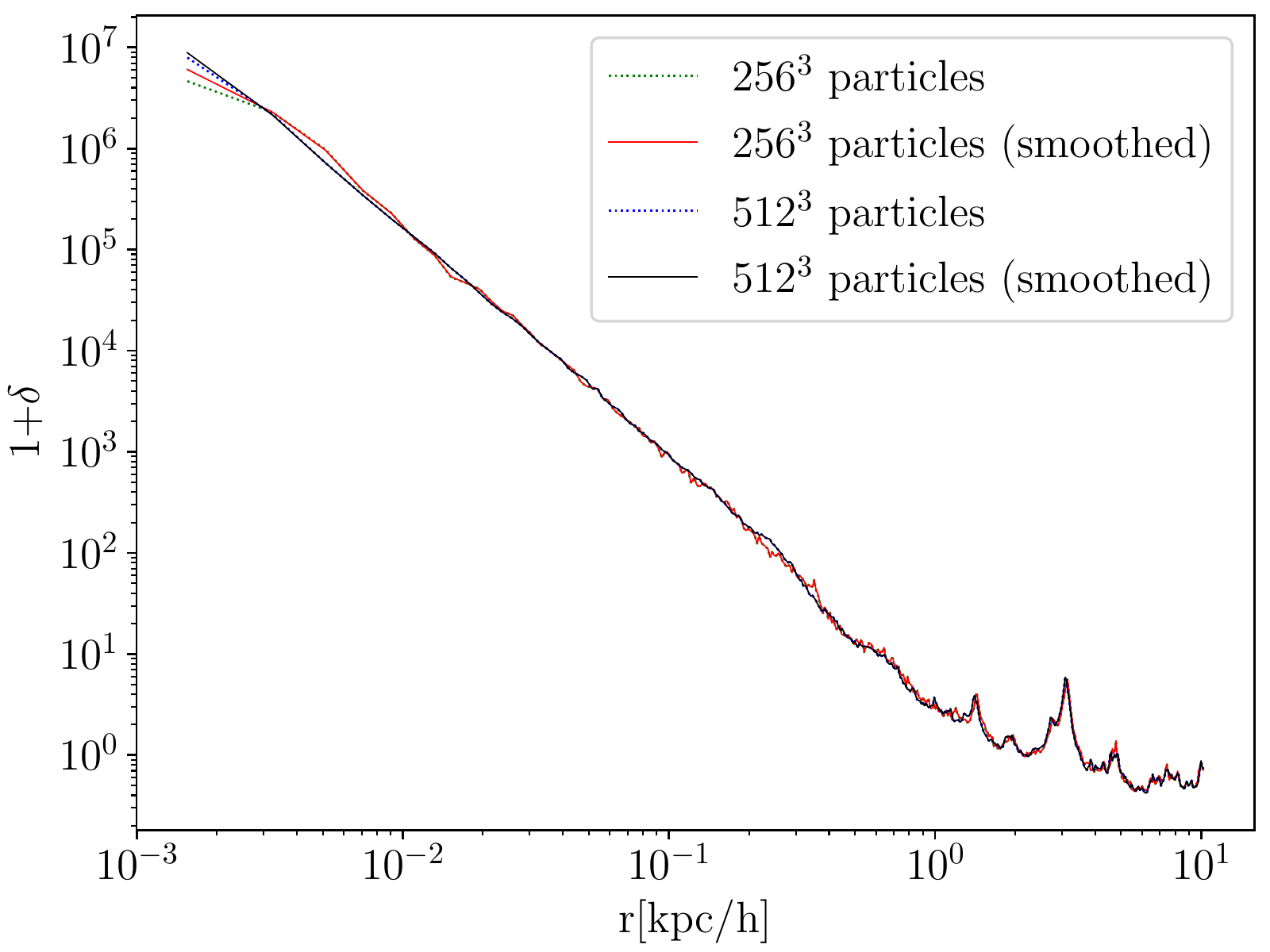}
\caption{\label{fig:sn15convergence} (Color online) The profile of the Gaussian-seed halo in the peak-to-background ratio $15$ simulation at $z=30$ for two different numbers of particles. We also show the effect of smoothing -- the profiles before applying smoothing are shown in dashed and the ones after are continuous. The smoothing we applied affects the shape of the profile even less than changing the resolution. The only noticeable difference is in the first bin.}
\end{figure}

\bibliographystyle{utcaps}
\bibliography{UCMH}

\end{document}